\documentclass[lettersize,journal]{IEEEtran}
\usepackage{amsmath,amsfonts,amssymb}
\usepackage{mathtools}
\usepackage{algorithm}
\usepackage{array}
\usepackage{booktabs}
\usepackage[font=small,labelfont=bf]{caption}
\usepackage{subcaption}
\usepackage{textcomp}
\usepackage{stfloats}
\usepackage{url}
\usepackage{verbatim}
\usepackage{graphicx}
\usepackage{svg}
\usepackage[hidelinks]{hyperref} 
\usepackage[acronym,shortcuts,nonumberlist]{glossaries} 
\makeglossaries
\newacronym{DBMC}{DBMC}{diffusion-based molecular communication}
\newacronym{TDMA}{TDMA}{time-division multiple access}
\newacronym{MDMA}{MDMA}{molecule-division multiple access}
\newacronym{NOMA}{NOMA}{non-orthogonal multiple access}
\newacronym{ADMA}{ADMA}{amplitude-division multiple access}
\newacronym{MA}{MA}{multiple access}
\newacronym{BEP}{BEP}{bit error probability}
\newacronym{TX}{TX}{transmitter}
\newacronym{RX}{RX}{receiver}
\newacronym{ISI}{ISI}{inter-symbol interference}
\newacronym{MAI}{MAI}{multiple-access interference}
\newacronym{SIC}{SIC}{successive interference cancellation}
\newacronym{OOK}{OOK}{on-off-keying}
\newacronym{MCS}{MCS}{Monte Carlo simulation}
\newacronym{SNR}{SNR}{signaling-molecule-to-noise ratio}
\newacronym{MC}{MC}{molecular communication}
\newacronym{IoBNT}{IoBNT}{Internet of Bio-Nano-Things}
\newacronym{BNM}{BNM}{Bio-Nano-Machine}
\newacronym{DBMC-NOMA}{DBMC-NOMA}{NOMA for DBMC networks}
\newacronym{UCA}{UCA}{uniform concentration assumption}
\newacronym{CRN}{CRN}{chemical reaction network}
\newacronym{ODE}{ODE}{ordinary differential equation}
\newacronym{SSA}{SSA}{stochastic simulation algorithm}
\newacronym{ChemSICal}{ChemSICal}{chemical reaction network for successive interference cancellation}
\newacronym{PDF}{PDF}{probability density function}
\newacronym{PMF}{PMF}{probability mass function}
\newacronym{RRC}{RRC}{reaction rate constant}
\newacronym{SCRN}{SCRN}{stochastic chemical reaction network}
\newacronym{DBMC-aNOMAly}{DBMC-aNOMAly}{asynchronous NOMA with pilot-symbol optimization protocol for DBMC}
\newacronym{EM}{EM}{electromagnetic}
\newacronym{IoT}{IoT}{Internet of Things}
\newacronym{MI}{MI}{mutual information}
\newacronym{WCAM}{WCAM}{worst-case-offset avoidance mechanism}
\newacronym{BO}{BO}{Bayesian optimization}
\newacronym{MCMC}{MCMC}{Markov chain Monte Carlo}
\newacronym{MH}{MH}{Metropolis-Hastings}
\newacronym{SA}{SA}{simulated annealing}
\newacronym{AM}{AM}{approximate majority}
\newacronym{PFBL}{PFBL}{positive feedback loop}
\newacronym{NFBL}{NFBL}{negative feedback loop}
\newacronym{CDF}{CDF}{cumulative density function}
\newacronym{CI}{CI}{confidence interval}  
\glsdisablehyper
\usepackage{cite}
\usepackage{todonotes}
\usepackage{lipsum}
\usepackage{siunitx}
\usepackage{algpseudocodex}
\usepackage{titlesec}

\usepackage{supertabular}
\newglossarystyle{acr-2col-narrow}{%
  \renewcommand*{\glsgroupheading}[1]{}
}

\titlespacing\section{0pt}{3pt plus 2pt minus 2pt}{4pt plus 2pt minus 2pt}
\titlespacing\subsection{0pt}{2pt plus 1pt minus 1pt}{2pt plus 1pt minus 1pt}

\begin{document}

\title{ChemSICal-Net: Timing-Controlled Chemical Reaction Network for Successive Interference Cancellation in Molecular Multiple Access

\thanks{The authors acknowledge the financial support by the German Federal Ministry of Research, Technology and Space (BMFTR) in the program of “Souverän. Digital. Vernetzt.”. Joint project 6G-life, project identification number: 16KISK002.}
\thanks{Alexander Wietfeld and Wolfgang Kellerer are with the Chair of Communication Networks, Technical University of Munich, 80333 Munich, Germany (e-mail: \{alexander.wietfeld, wolfgang.kellerer\}@tum.de). Oguz Turgut and Eneritz Somoza Rodríguez were with the Chair of Communication Networks, Technical University of Munich, 80333 Munich, Germany, at the time this work was conducted.}}

\author{Alexander Wietfeld,~\IEEEmembership{Graduate Student Member,~IEEE}, Oguz Turgut, Eneritz Somoza Rodr\'iguez,\\ Wolfgang Kellerer,~\IEEEmembership{Fellow,~IEEE}
}



\maketitle

\begin{abstract}
Molecular communication (MC) networks are envisioned to enable synthetic information exchange between nanoscale biological entities. For many algorithm proposals in the MC research field, the question of implementation at nanoscales and in biological environments remains open. Chemical reaction networks (CRNs) provide a natural framework to model computing processes in biological systems, while detailed simulations capture realistic stochastic effects. In this work, we present ChemSICal-Net, a comprehensive CRN simulation model of a chemical receiver implementing successive interference cancellation (SIC) to differentiate messages from multiple transmitters. We present the structure of the SIC algorithm in the form of basic chemical building blocks and incorporate clocked timing control by a chemical oscillator. We propose an adaptive Bayesian optimization (BO) scheme with a Gaussian process surrogate to find appropriate values for the reaction rate constants and the initial concentrations and show that it outperforms baseline methods from related work based on a fair computational cost metric. Then, the performance of the ChemSICal-Net framework is evaluated stochastically across a range of clock speeds and in different configurations focusing on communication system metrics such as detection accuracy and decision time. Our results highlight that the timing via a chemical clock can improve the detection accuracy by a factor of 2 in scenarios with shorter decision times, which underlines how the trade-off between decision time and detection probability can shape CRN design choices. The BO scheme is shown to reliably optimize parameters for different configurations by approximately one order of magnitude compared to the non-optimized case. Our system reveals the need for a multi-scale approach with external BO and stochastic simulation of molecular reaction dynamics for communication-metric-focused system design.
\end{abstract}

\begin{IEEEkeywords}
Molecular communication, chemical reaction networks, NOMA, successive interference cancellation, chemical oscillator, Bayesian optimization
\end{IEEEkeywords}

\section{Introduction}\label{sec:introduction}

\IEEEPARstart{M}{olecular} communication (MC) is defined as the information transfer using molecules~\cite{farsad_comprehensive_2016}. 
As a biocompatible and energy-efficient alternative to classical electromagnetic signals, \ac{DBMC}~\cite{jamaliChannelModelingDiffusive2019} will play a major role in a future \ac{IoBNT}, connecting bio-nano nodes inside the human body~\cite{akyildizInternetBioNanoThings2015}. Recent advances in bio-nanotechnology and micro-robotics have shown that monitoring and targeted interventions have the potential to enable novel medical use cases~\cite{landersClinicallyReadyMagnetic2025a}, but still require external tools and human operation. 

Biological devices at small scales, so-called \acp{BNM}, will offer severely limited communication capabilities and computing resources~\cite{elaniVesiclebasedArtificialCells2014}. To realize the \ac{IoBNT} vision, autonomous networked systems and novel computing approaches are needed to connect multiple \acp{BNM} and implement more complex decision mechanisms. As a result, multi-user communication becomes an important reference case, for example, where multiple \acp{TX} report to a shared \ac{RX} for data gathering~\cite{shitiriTDMABasedDataGathering2021}. Interference management and avoidance is crucial when multiple transmissions need to be coordinated. Therefore, molecular \ac{MA} has received significant attention, including orthogonal and non-orthogonal approaches relying on procedures and algorithms such as \ac{SIC}~\cite{shitiriTDMABasedDataGathering2021, wangPracticalScalableMolecular2023, wietfeldDBMCNOMAEvaluatingNOMA2024}. However, most \ac{MA} schemes for \ac{DBMC} are evaluated at the level of abstract \ac{RX} models and typically assume digital processing resources that do not exist in \acp{BNM} to execute algorithms.

This gap motivates research into chemical implementations of signal processing and computing. \Acp{CRN} are combinations of chemical reactions that by specific coupling of reactants and products of different reactions enable the implementation of simple mathematical and logical functions~\cite{vasicCRNMolecularProgramming2020}. As a consequence, they have become a promising substrate for \ac{TX} and \ac{RX} functionality in \ac{DBMC} networks both in experiments~\cite{biChemicalReactionsBasedMicrofluidic2020, walterRealtimeSignalProcessing2023}, as well as simulation studies~\cite{heinleinClosingImplementationGap2024}. Recent work suggests that \acp{CRN} provide a credible path toward algorithm implementation without relying on electronic and digital components in biological environments. Therefore, proposed algorithms must be robust under the influence of chemical dynamics and inherent stochasticity of \acp{CRN}~\cite{heinleinClosingImplementationGap2024}.

In this context, we introduced ChemSICal as a stochastic \ac{CRN} simulation model of \ac{SIC} for \ac{NOMA} in \ac{DBMC} networks~\cite{wietfeldChemSICalEvaluatingStochastic2025}, performed a preliminary  evaluation and investigated how chemical parameters like \acp{RRC} influenced performance. Building on this foundation, the goal of this paper is to advance from a standalone \ac{CRN} \ac{SIC} implementation toward a full \ac{RX} system that adds two crucial missing aspects: explicit timing control of multi-stage processing using a chemical oscillator, and systematic optimization in high-dimensional \ac{CRN} parameter spaces.

\begin{figure*}
    \centering
    \includegraphics[width=0.55\linewidth]{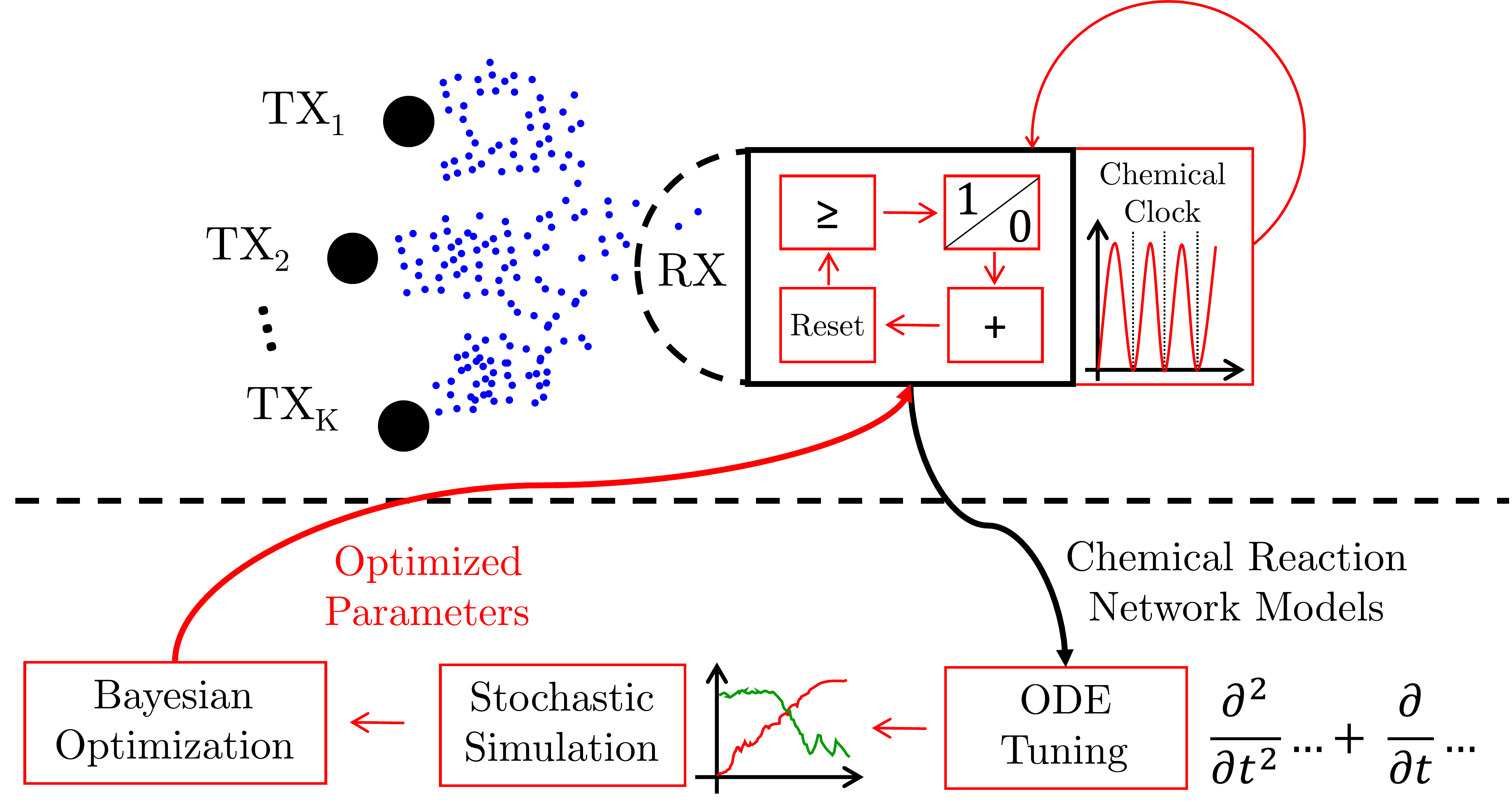}
    \caption{Schematic representation of the ChemSICal-Net framework, showing the \ac{DBMC} channel with multiple \aclp{TX}, and the \acl{RX} with block-based chemical computing operations and chemical clock for timing. We showcase the multi-scale design loop: The system can be expressed as \acf{ODE}, which can be used for initial parameter tuning. Then, comprehensive stochastic simulations form the input for our proposed \acf{BO} scheme.}
    \label{fig:teaser}
\end{figure*}

We address how a chemical \ac{SIC} \ac{RX} can be realized as a \ac{CRN} connected to a \ac{DBMC} channel with dynamic inputs~\cite{wietfeldChemSICalEvaluatingStochastic2025} and evaluate the system thoroughly with \ac{ODE} solvers and stochastic simulations. Further, we ask how a chemical timing process can be integrated into the multi-\ac{TX} \ac{SIC} algorithm to gate different stages and make the decision and readout timing an explicit design parameter. We investigate the conditions under which the chemical timing can improve performance via added control, or when the added reaction complexity can degrade performance compared to the non-timed version. Finally, we investigate how \ac{BO}, which has a proven track record for usage in complex chemical systems~\cite{shieldsBayesianReactionOptimization2021, braconiBayesianOptimizationValuable2023, guoBayesianOptimizationChemical2023}, can be adapted to tune constrained stochastic \acp{CRN} with communication-centric metrics such as error probability. The resulting \textit{ChemSICal-Net} framework targets an end-to-end view of a chemical \ac{SIC} \ac{RX} as a \ac{DBMC} system component and includes modular proof-of-concept components toward reuse through a reset mechanism and scaling beyond two simultaneous \acp{TX}. An overview of the proposed concept can be seen in Figure~\ref{fig:teaser}.

\subsection{State of the Art}

In the following, the state-of-the-art in relevant research areas will be summarized and relevant gaps highlighted. This includes \ac{MA} for \ac{DBMC} networks, \acp{CRN} for modeling chemical algorithms, using chemical oscillators to control \ac{CRN} algorithms, and \ac{BO} for chemical reaction tuning.

\subsubsection{Multiple access in DBMC networks}
\Ac{MA} schemes have been explored extensively in \ac{DBMC} literature as one of the first important steps towards multi-node networks. One main category covers orthogonal approaches such as \ac{TDMA}~\cite{shitiriTDMABasedDataGathering2021}, where each \ac{TX} is assigned a specific time slot, or \ac{MDMA}~\cite{chenResourceAllocationMultiuser2021}, which involves different molecule types per \ac{TX} that can be chemically differentiated at the \ac{RX}. Advanced approaches exist, for example, using molecule mixtures for \ac{MDMA}, inspired by olfactory systems~\cite{jamaliOlfactioninspiredMCsMolecule2023}. While orthogonal approaches inherently reduce interference, they incur different types of overhead, such as strict timing requirements and a lack of simultaneous transmissions for \ac{TDMA}, or increased chemical complexity through required compatibility of \ac{TX} and \ac{RX} with additional molecule species for \ac{MDMA}. \Ac{NOMA}-inspired approaches have been explored recently~\cite{wangPracticalScalableMolecular2023, wietfeldDBMCNOMAEvaluatingNOMA2024} to allow for simultaneous transmission from multiple \acp{TX} with the same molecule type. Similar concepts have been shown to be successful at increasing capacity in classical communication networks~\cite{saitoNOMA2013}. In the most common form, power-domain \ac{NOMA}, the \ac{RX} experiences additional interference from simultaneously transmitting \acp{TX}, which can be mitigated using a \ac{SIC} algorithm based on differences in received signal strength~\cite{wietfeldErrorProbabilityOptimization2024c}.

The most prevalent style of evaluation for \ac{DBMC} \ac{MA} literature remains based on abstract \ac{RX} models. This has revealed important insights about performance and design trade-offs between different \ac{MA} schemes~\cite{wietfeldDBMCNOMAEvaluatingNOMA2024}, but the realization of the necessary algorithms, for example, for synchronization, coding, or \ac{SIC} in bio-nano environments is still largely unaddressed. Despite the maturity of the research area, this leaves an open gap regarding the realistic modeling of \ac{MA} scheme implementations in \ac{DBMC} networks without generic digital computing capabilities.

\subsubsection{CRNs for computing in DBMC networks}

\begin{figure}[t]
    \centering
    \includegraphics[width=0.65\linewidth]{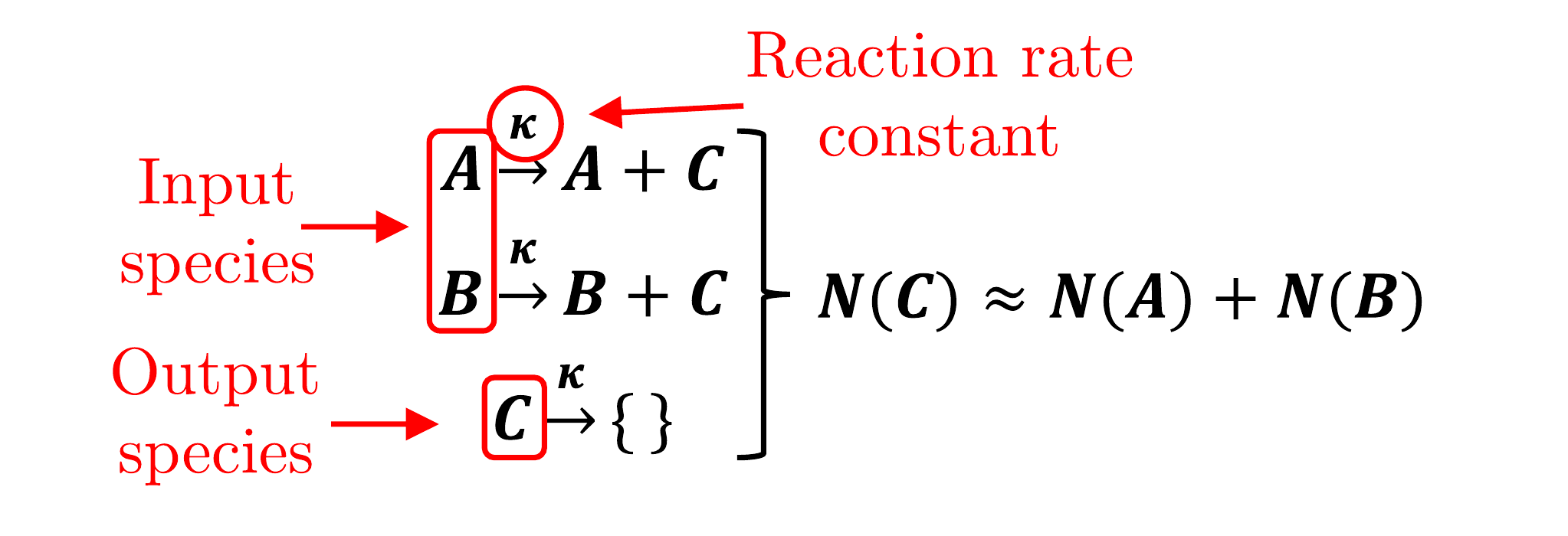}
    \caption{Example of a simple \ac{CRN} computing the sum of the molecule counts of two input species A and B as the molecule count of a third chemical species C.}
    \label{fig:crn_example}
\end{figure}

A chemical reaction is defined by molecule species inputs (reactants), molecule species outputs (products) and the speed of the reaction between them (\acp{RRC}). Figure~\ref{fig:crn_example} shows an example of a \ac{CRN}, in which multiple chemical reactions interact to implement a simple mathematical addition function. Research into molecular programming frameworks has shown that \acp{CRN} offer an appropriate modeling substrate for chemical computing using reaction blocks that enable further operations like comparisons or multiplication~\cite{vasicCRNMolecularProgramming2020}. Heinlein et al. have recently proposed the first \ac{CRN}-based \ac{DBMC} \ac{RX} that includes symbol detection and synchronization for single-link communication~\cite{heinleinClosingImplementationGap2024}. Other related work has considered \ac{CRN}-like building blocks and precise interplay of reaction volumes to create simple neural network structures in the chemical domain~\cite{angerbauerMolecularNanoNeural2024}. In addition, recent work has highlighted the specific role of DNA-based reactions for the implementation of  programmable molecular protocols~\cite{lauCommunicationProtocolsMolecular2024}, as well as experimental \ac{DBMC} systems~\cite{wietfeldModelingMicrofluidicInteraction2025b}.

These papers establish that chemical computation and \ac{DBMC}-node signal processing using \acp{CRN} is possible and promising in principle, and experimental work has shown artificial cells can act as chemical reactors containing \ac{CRN} processes~\cite{elaniVesiclebasedArtificialCells2014}. However, remaining challenges are the integration of \ac{CRN}-based algorithms into multi-user networks, the connection with a dynamic \ac{DBMC} channel, and the evaluation using communication metrics and stochastic simulations. While some works address \acp{CRN} using stochastic simulations~\cite{heinleinClosingImplementationGap2024}, they lack the multi-user network aspect, and other consider more complex protocols~\cite{lauCommunicationProtocolsMolecular2024}, but do not integrate the system into a diffusion channel and communication-focused evaluation.

\subsubsection{Chemical oscillators for timing control}

Using chemical oscillators to control the flow of \acp{CRN} has long been a topic of research. Prior work has already proposed clocked schemes that enable sequential computation, where oscillatory signals gate certain reaction blocks one by one~\cite{jiangSynchronousSequentialComputation2011, jiangSynthesisFlowDigital2010}. While early work makes use of simplified oscillator models, recent work has specifically framed chemical oscillators as clock signals to be optimized for chemical computation and tuned parameters and oscillator structure to achieve properties such as symmetry~\cite{shiDesignUniversalChemical2022}. While the previous papers target \textit{toy model} oscillators, which do not exist in natural systems, the biological literature provides multiple oscillator types, including cell-cycle~\cite{cardelliCellCycleSwitch2012}, gene-regulatory~\cite{tsaiRobustTunableBiological2008}, or phosphorylation-driven~\cite{Mixed_mechs_of_multi-site_phosphor} oscillations. These models represent attractive candidates, when it comes to biophysical plausibility.

The research gap consists of the use of chemical oscillators specifically for \ac{DBMC} purposes with a systematic comparison of oscillator types based on clock signal metrics. Further, there is little work that connects oscillator parameter choice regarding, for example, clock frequency with communication metrics such as error probability. Lastly, most literature disregards stochastic fluctuations~\cite{jiangClockSignalGeneration2025}, which have a major effect on probabilistic communication performance.

\subsubsection{Bayesian optimization for stochastic CRNs}

\Ac{CRN} design typically involves high-dimensional parameter spaces, non-linear dynamics, and high-effort stochastic simulations, especially when the system is supposed to operate at low molecule counts and in single-cell environments~\cite{heinleinClosingImplementationGap2024}. Manual tuning and small-scale parameter sweeps, as have been applied in existing work on \acp{CRN} for \ac{DBMC}~\cite{heinleinClosingImplementationGap2024, jiangClockSignalGeneration2025, wietfeldChemSICalEvaluatingStochastic2025}, can provide a first intuition but do not scale to complex protocols or allow for comprehensive performance comparison. 
Previous work in \ac{CRN} literature regarding simulation models has explored different approaches to systematic optimization. Vasi\'c et al.~\cite{vasicProgrammingTrainingRateindependent2022} have investigated a special class of rate-independent \acp{CRN}, which can be optimized purely with regards to stoichiometry, i.e. the numbers of molecules, as opposed to the \acp{RRC}. These networks are mostly limited to implementing piecewise-linear functions, and so the authors show how they could be useful in building neural networks with rectified linear unit function blocks. 

Since complex stochastic \ac{CRN} models are expensive to evaluate and usually do not allow for analytical or explicit gradient-based optimization, efficient sampling-based global optimization solutions for black-box functions are commonly employed. Murphy et al.~\cite{murphySynthesizingTuningStochastic2018b}  consider synthesizing \acp{CRN} for certain functions and propose a method of optimizing the reaction parameters through a \ac{MCMC} method with a \ac{MH} algorithm to estimate the posterior distribution of the parameter set from a model~\cite{robertMonteCarloStatistical2004}. The method tries to bias the search towards promising regions to speed up optimization. 
\Ac{SA} is another stochastic search scheme that explores the parameter space based on a temperature schedule allowing it to escape local minima early and later focus on gradually refining the objective as the temperature decreases~\cite{kalivasAdaptionSimulatedAnnealing1995}. Due to its general applicability to expensive objective functions, \ac{SA} has been successfully applied to \ac{CRN} optimization problems~\cite{ashyraliyevSystemsBiologyParameter2009}.
Lastly, \acf{BO} tries to build a probabilistic surrogate function of the expensive black-box model and applies an acquisition rule for the next sampling point that balances exploration and exploitation~\cite{guoBayesianOptimizationChemical2023}. \Ac{BO} has recently become a practical tool for efficient optimization of expensive and noisy objectives, including real-world reaction optimization under experimental constraints~\cite{shieldsBayesianReactionOptimization2021, braconiBayesianOptimizationValuable2023, guoBayesianOptimizationChemical2023}. Therefore, it appears as a promising approach for the optimization of stochastic simulation models for \acp{CRN}.

What remains largely open is the evaluation of \ac{BO} against other optimization methods as a system-level tuning mechanism for constrained stochastic \acp{CRN} in the context of \ac{DBMC} objectives. Instead of chemical yield or general behavioral specifications, the objective should target metrics such as error probability or decision time. This leads to a natural multi-scale approach where the computationally heavy optimization is performed externally during design, while the deployed \ac{CRN}-based \ac{DBMC} system is self-contained within its constrained environment.

\subsubsection{Summary of gaps}

Across these bodies of work, a clear missing intersection emerges. Multi-user \ac{DBMC} \ac{MA} is well studied at the system level, and chemical \ac{RX} pipelines and timing primitives exist in related contexts. However, end-to-end designs that implement multi-user \ac{SIC} as a stochastic \ac{CRN}, integrate an explicit chemical timing primitive, and systematically tune constrained parameters against \ac{DBMC} communication objectives remain scarce. 

In~\cite{wietfeldChemSICalEvaluatingStochastic2025}, we presented ChemSICal as the first explicit effort to map \ac{SIC} into a stochastic \ac{CRN} and evaluated it under \ac{DBMC} channel conditions. The work demonstrated that realistic chemical models can deviate substantially from the idealized \ac{SIC} assumptions due to stochastic effects and reaction competition. At the same time, it revealed major gaps in the system integration. First, multi-stage processing in \acp{CRN} is logically sequential but physically concurrent, which makes it difficult to enforce stage boundaries and validate early-stage decisions in time. Second, it highlights the high parameter dimensionality and sensitivity that makes it hard to tune the \ac{CRN} parameters manually or analytically. Lastly, the ChemSICal system is limited to 2 \acp{TX} and does not address issues with repeated use of the \ac{CRN} via a reset or state-clearing mechanism.

This motivates ChemSICal-Net as a step toward network-ready chemical signal processing.

\subsection{Contributions and extension of previous work}

This work builds upon our previous work in~\cite{wietfeldChemSICalEvaluatingStochastic2025} and the state of the art as described above by making several contributions:

\begin{enumerate}
    \item \textbf{ChemSICal-Net: an \ac{MA}-oriented chemical \ac{SIC} \ac{RX} framework}: We present an end-to-end \ac{CRN} design that implements \ac{SIC}-based multi-user reception and interfaces with \ac{DBMC} channel observations. We deepen our initial analysis in~\cite{wietfeldChemSICalEvaluatingStochastic2025} via comprehensive multi-scale evaluation using deterministic \ac{ODE} analysis for coarse tuning and stochastic simulations for detailed performance analysis, quantifying where and how chemical models deviate from idealized assumptions.
    \item \textbf{Clocked execution with chemical oscillator control}: Towards system integration, we incorporate a chemical timing primitive to control processing stages and make decision time a controllable design parameter. To that end, we systematically compare multiple relevant candidate oscillators~\cite{cardelliCellCycleSwitch2012, Mixed_mechs_of_multi-site_phosphor, tsaiRobustTunableBiological2008} via stochastic simulations and frequency-domain analysis, and identify a dual-site phosphorylation oscillator~\cite{Mixed_mechs_of_multi-site_phosphor} as the most stable and easily parametrizable.
    \item \textbf{System-level trade-off evaluation}: By comparing a clocked and non-clocked variant and evaluating the effect of decision time on the error probability under \ac{RRC} constraints, we highlight how timing control can support \ac{CRN} execution particularly in higher-frequency decision regimes, but that the lower-complexity non-clocked variant can perform better in the low-decision-frequency regime.
    \item \textbf{Adaptive \ac{BO} scheme for constrained stochastic \ac{CRN} tuning}: We propose a \ac{BO}-based optimization approach with adaptive stochastic replication, particularly suited for complex stochastic systems with limited analytical tractability and in line with state-of-the-art methods for experimental chemical reaction optimization. We compare the optimization performance efficiency of the proposed scheme with related optimization schemes and show that it exhibits superior results. Subsequently, we integrate the \ac{BO} scheme into the ChemSICal-Net framework by targeting communication system objectives.
    \item \textbf{Extension via reuse and scaling}: Lastly, we consider proof-of-concept extensions of the ChemSICal-Net system. We propose a simple reset mechanism and perform a preliminary evaluation of its effectiveness. Further, we outline a general path to adding more \acp{TX} to the system and discuss challenges with scalability.
\end{enumerate}

The remainder of this paper is structured as follows:
In Section~\ref{sec:system_model}, the physical channel model, communication system, and idealized \ac{SIC} algorithm for \ac{NOMA} are introduced. In Section~\ref{sec:oscillators}, we introduce three common and relevant models for chemical oscillation, present evaluation results and choose the most stable option for our system. Then, Section~\ref{sec:crn_design} presents the structural design of the ChemSICal-Net framework, including integration of the oscillator, as well as reset and \ac{TX} scaling, while Section~\ref{sec:simulation} explains the metrics and simulation methods under which we will conduct the evaluation. In Section~\ref{sec:optimization}, we showcase the proposed \ac{BO} scheme and conduct example evaluations against two benchmark schemes before presenting comprehensive simulation results in Section~\ref{sec:evaluation}. We compare different variants of the ChemSICal-Net system, and quantify the impact of timing control and optimization on the communication performance. Lastly, Section~\ref{sec:conclusion} outlines the conclusions. Detailed parameter tables and chemical reaction definitions are given in the Appendix as supplementary material.

\section{Scenario and System Model}\label{sec:system_model}

\begin{figure}[tb]
    \centering
    \includegraphics[width=0.75\linewidth]{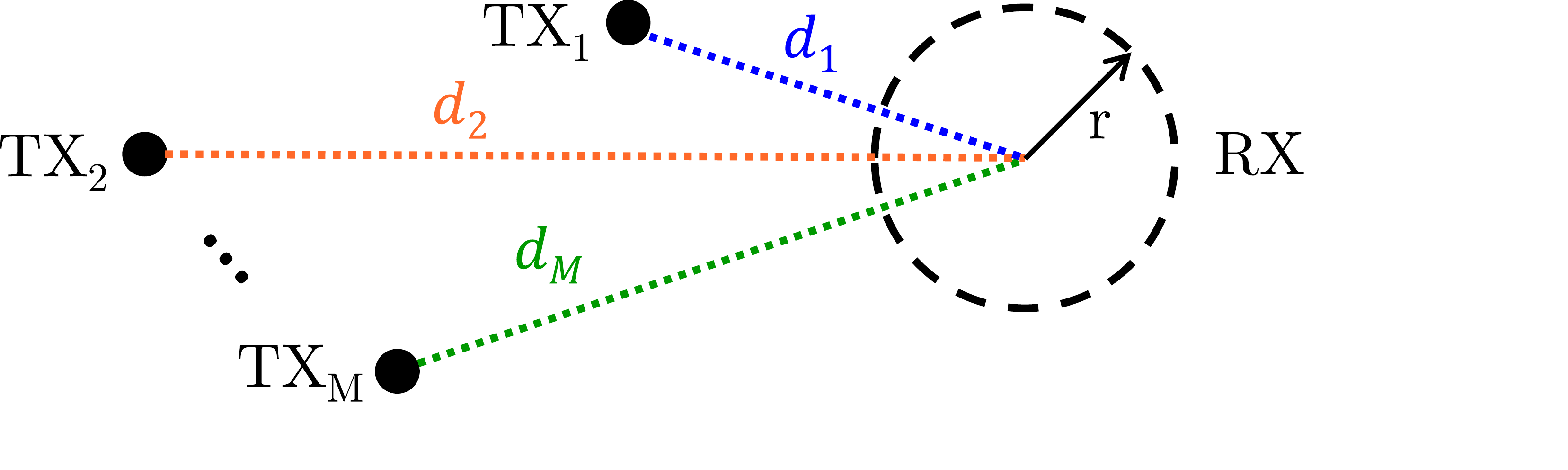}
    \caption{\ac{DBMC} scenario with $M$ point \acp{TX} at distances $d_1$, $d_2$ ... $d_M$ from a spherical \ac{RX}.}
    \label{fig:scenario}
\end{figure}

We consider a \ac{DBMC} network with $M$ point \acp{TX}, \ac{TX}$_1$,~$\dots$, \ac{TX}$_M$ located at distances $d_1\leq d_2 \leq \dots \leq d_M$ from a single spherical \ac{RX} of radius $r$, as illustrated in Figure~\ref{fig:scenario}.
The \ac{RX} is assumed to be a passive observer of the instantaneous number of molecules inside its volume $V_\mathrm{RX} = \frac{4}{3}\pi r^3$ over time, forming the received signal $n_\mathrm{RX}(t)$.
Molecules propagate in an unbounded three-dimensional environment via Brownian motion with diffusion coefficient $D$, and we adopt the \ac{UCA} in the regime $r < 0.15 d_i$~\cite{jamaliChannelModelingDiffusive2019} to apply the closed-form impulse response model.
The \acp{TX} are assumed to be one-dimensional points capable of instantaneously emitting molecules.
Thereby, for TX$_i$ emitting $N_{\mathrm{TX}}$ molecules at time $t=0$ and distance $d_i$ from the \ac{RX}, the average contribution to the received signal over time $t$ is given by~\cite{jamaliChannelModelingDiffusive2019}
    \begin{equation}
     \lambda_i(t) = \frac{N_{\mathrm{TX}}V_\mathrm{RX}}{\left(4\pi Dt\right)^\frac{3}{2}}\exp \left(-\frac{d_i^2}{4Dt}\right)\ \mathrm{for}\ t\geq0.  
    \end{equation}

\subsection{Communication System}\label{subsec:communication_system}

Each \ac{TX} employs \ac{OOK} to transmit information with equiprobable symbols $s_i\in\{0,1\}$ by emitting either $N_\mathrm{TX}$ molecules for a '1' or zero molecules for a '0' at the beginning of each symbol interval.
We assume independent molecule behavior and model the received signal contribution from each \ac{TX} as a Poisson-distributed random variable with mean $\lambda_i (t)$. This assumption is valid for sufficiently large $N_\mathrm{TX}$ and $\lambda_i(t)\ll N_\mathrm{TX}$~\cite{torresgomezAgeInformationMolecular2022}, conditions which hold for our parameter choices later, see Table~\ref{tab:sys_params_main}.
To keep the chemical evaluation focused on the \ac{RX}-side, we assume a baseline scenario that is fully synchronized, such that all signals arrive at the \ac{RX} simultaneously, and we select a sufficiently large symbol period such that \ac{ISI} is negligible. We assume peak sampling at the \ac{RX}, such that each symbol interval is characterized by a single sample. Specifically, the \ac{RX} takes one sample at the peak time $t_\mathrm{p}$ of the received signal as the sampled number of molecules $n_\mathrm{s} = n_\mathrm{RX}(t_\mathrm{p})$, which is the only input used by the subsequent \ac{SIC} detection algorithm~\cite{wietfeldChemSICalEvaluatingStochastic2025}.
As defined by the \ac{DBMC-NOMA} scheme~\cite{wietfeldDBMCNOMAEvaluatingNOMA2024}, all \acp{TX} transmit simultaneously using the same molecule type. Since the sum of multiple Poisson variables is itself Poisson distributed, the sample $n_\mathrm{s}$ follows a Poisson distribution, i.e.
    \begin{equation}
        n_\mathrm{s} \sim \mathcal{P}\left( \boldsymbol{s}\cdot\boldsymbol{\lambda}\right).
    \end{equation}
with the transmitted symbol pattern $\boldsymbol{s}=[s_1,\dots,s_M]\in \{0,1\}^M$ and the vector of corresponding received averaged $\boldsymbol{\lambda} = [\lambda_1, \dots, \lambda_M]$.
For the general multi-user case with $M$ \acp{TX} and equiprobable \ac{OOK} symbols, the induced distribution over $n_\mathrm{s}$ can be expressed as a mixture of $2^M$ Poisson components, resulting in
    \begin{equation}\label{eq:input_pmf_K}
    p_{n_\mathrm{s}}[n]
    = \frac{1}{2^M}\sum_{\boldsymbol{s}\in\{0,1\}^M}
    \mathcal{P}_\mathrm{PMF}\left(n;\boldsymbol{s}\cdot\boldsymbol{\lambda}\right),
    \end{equation}
where $\mathcal{P}_\mathrm{PMF}(n;\lambda) = \lambda^n\frac{e^{-\lambda}}{n!}$ denotes the \ac{PMF} of a Poisson distribution evaluated at $n$ with mean $\lambda$.
For example, for the two-user baseline ($K=2$) utilized through most of the paper, the induced distribution over $n_\mathrm{s}$ is a mixture of 4 Poisson components with peaks at $n=0$, $\lambda_1$, $\lambda_2$, and $\lambda_1+\lambda_2$, respectively.

\subsection{NOMA and Successive Interference Cancellation}\label{subsec:noma_sic}

\begin{figure}[tb]
    \centering
    \includegraphics[width=0.6\linewidth]{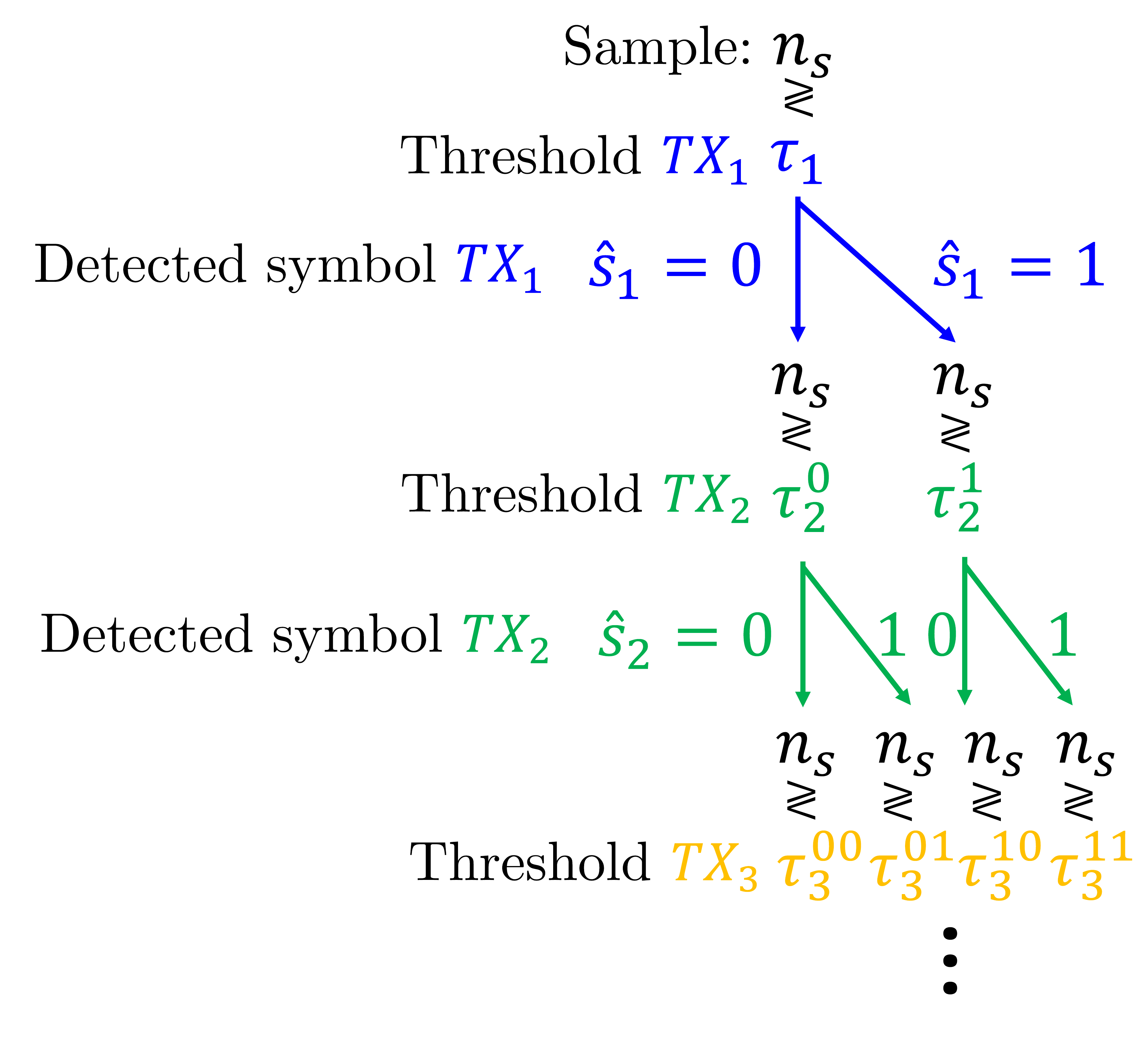}
    \caption{Simplified threshold-based \ac{SIC} algorithm for \ac{DBMC}, adapted from~\cite{wietfeldErrorProbabilityOptimization2024c}.}
    \label{fig:sic}
\end{figure}

By utilizing \ac{NOMA}, \acp{TX} transmit simultaneously and with the same molecule type. To deal with the resulting \ac{MAI}, \acf{SIC} is employed as the \ac{RX}-side algorithm that separates the transmitted symbols and assigns them to each \ac{TX} based on differences in the expected received amplitude~\cite{wietfeldDBMCNOMAEvaluatingNOMA2024}, for example due to differing distances $d_i$.
In the case of single-sample peak detection, \ac{SIC} can be described as a binary-tree threshold procedure, as depicted in Figure~\ref{fig:sic}, where the same $n_\mathrm{s}$ is compared multiple times with a threshold set that depends on previously detected symbols~\cite{wietfeldErrorProbabilityOptimization2024c}. We opt for this \ac{SIC} description for clarity and extensibility.
In the two-user case ($K=2$), \ac{SIC} uses a first-stage threshold $\tau_1$ to detect the symbol $\hat{s}_1$ for TX$_1$, and then selects $\tau_2^{\hat{s}_1} \in \{\tau_2^0, \tau_2^1\}$ for TX$_2$ detection.
For $K>2$, the threshold set generalizes to $\tau_i^{\boldsymbol{\hat{s}}_{i-1}}$ with $\boldsymbol{\hat{s}}_{i-1} = [\hat{s}_1, \dots, \hat{s}_{i-1}]$, yielding a full binary decision tree with $2^{i-1}$ thresholds at stage $i$.
The detection procedure for a given threshold produces the detected symbol $\hat{s}_i$ with 
    \begin{equation}
        \hat{s}_i = 
        \begin{cases}
            1 \qquad n_\mathrm{s}\geq \tau_i^{\boldsymbol{\hat{s}}_{i-1}},\\
            0 \qquad \mathrm{otherwise}.
        \end{cases}
    \end{equation}
For more details, we refer to our prior work, which provides an analytical error probability derivation for \ac{DBMC-NOMA} as well as optimization protocols~\cite{wietfeldDBMCNOMAEvaluatingNOMA2024, wietfeldDBMCaNOMAlyAsynchronousNOMA2025}.

\section{Comparison of Chemical Oscillators}\label{sec:oscillators}

A key limitation of the baseline ChemSICal design, as discussed in Section~\ref{sec:introduction}, is that all reactions execute simultaneously and correct \ac{SIC} behavior depends on the relative differences in \acp{RRC} resulting in an implicit reaction results ordering. Faster reactions produce results, which can still influence the slower reactions. The assumption of concurrent reactions is generally valid, as future \ac{IoBNT} nodes might only consist of a single synthetic cell acting as a reaction volume~\cite{elaniVesiclebasedArtificialCells2014}. However, in ChemSICal-Net, we introduce an explicit chemical clock to provide controlled time windows in which subsets of the reaction blocks are executed. The goal of this extension is to enable explicit speed-accuracy tradeoffs and avoid relying on high-precision \ac{RRC} separation across multiple orders of magnitude. On the other hand, adding an oscillating circuit will add complexity and noisy variations to the system, which could impact the stochastic performance negatively.
To control reaction blocks using a chemical clock, an oscillating species is incorporated into the associated reactions, so that corresponding reactions are gated by the presence of this oscillating species. Reaction blocks can then be ordered by choosing complementary non-overlapping oscillating species and using them to control subsequent reaction block steps~\cite{vasicCRNMolecularProgramming2020}. For this to work reliably, the frequency and amplitude of the oscillating species need to be stable across many oscillation periods.

In this section, we describe and evaluate a set of candidate oscillators as timing sources. We choose three representative models for detailed analysis using a \ac{MCS} approach with repeated stochastic simulations and a frequency-domain stability analysis. The simulations are conducted using the Gillespy2 package in Python~\cite{abelGillesPyPythonPackage2016}. The chosen models cover a range of structural mechanisms and biological origins and offer simple parametrization.
For each oscillator, we generate 500 trajectories over a fixed observation window of $T_\mathrm{obs} = 2000$ with 4001 sampling points. $T_\mathrm{obs}$ is given in normalized time relative to the \acp{RRC}. The observation time ensures a stable frequency estimation during the fast Fourier transform processing of $\Delta f = 1/T_\mathrm{obs} = 5\cdot10^{-4}$. This allows us to precisely determine the fundamental frequency $f_1$ and its amplitude $H_1$ and their statistical distributions. The parameters of each model are tuned to produce a target fundamental frequency of $f_1 = 0.1$, corresponding to about 200 oscillation periods per trajectory and a total of $10^5$ oscillation periods per model. Additionally, the peak number of molecules of the oscillating species is also tuned to be approximately 2000 for all models. 

\begin{figure}
    \centering
    \includegraphics[width=0.75\linewidth]{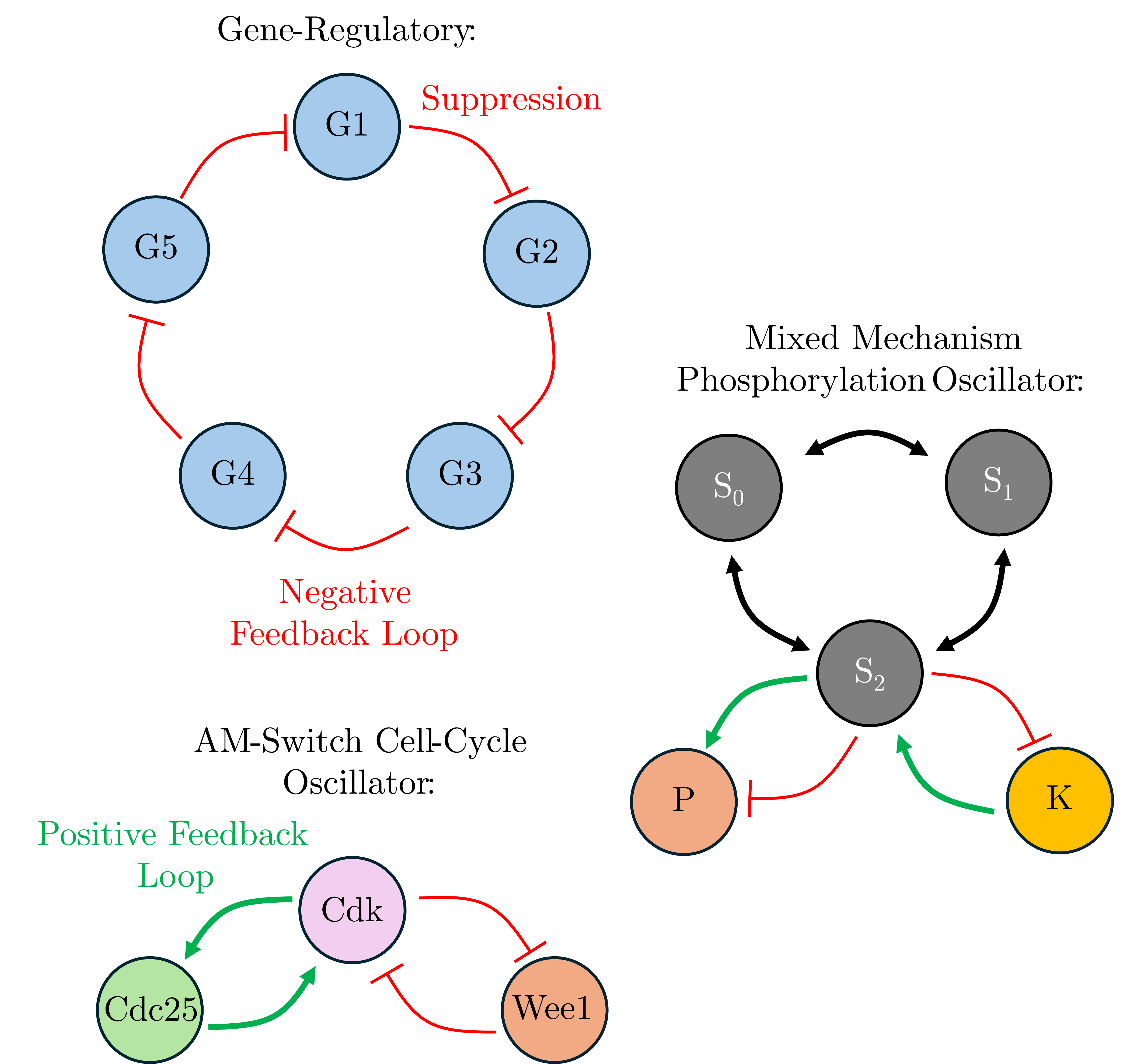}
    \caption{Simplified structural diagram of the gene-regulatory \textit{Pentilator}, the cell-cycle-based \acf{AM} switch oscillator, and a mixed mechanism phosphorylation oscillator. Circles represent molecules/genes, arrows represent relationships such as repression and amplification.}
    \label{fig:osc_diagrams}
\end{figure}

\begin{figure}
    \centering
    \includegraphics[width=0.31\linewidth]{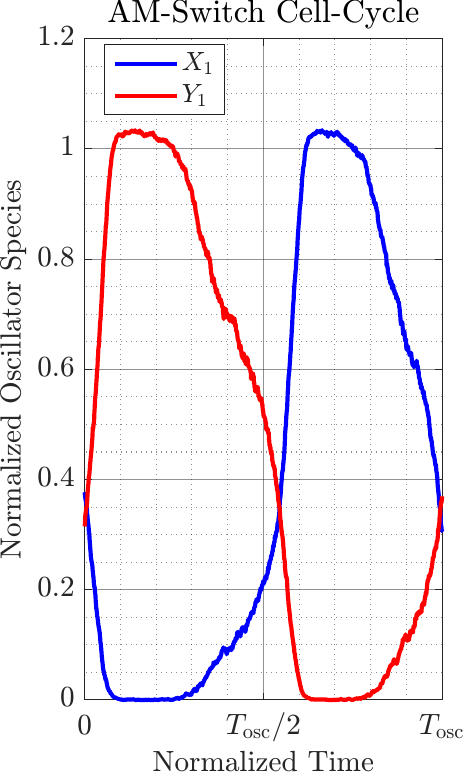}\includegraphics[width=0.31\linewidth]{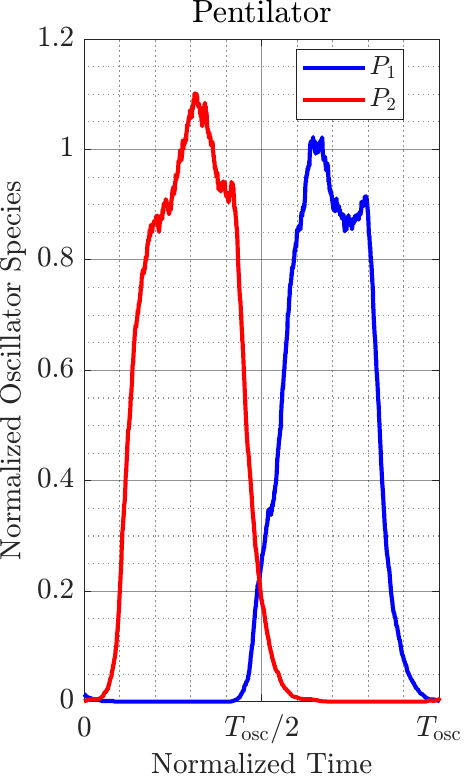}\includegraphics[width=0.31\linewidth]{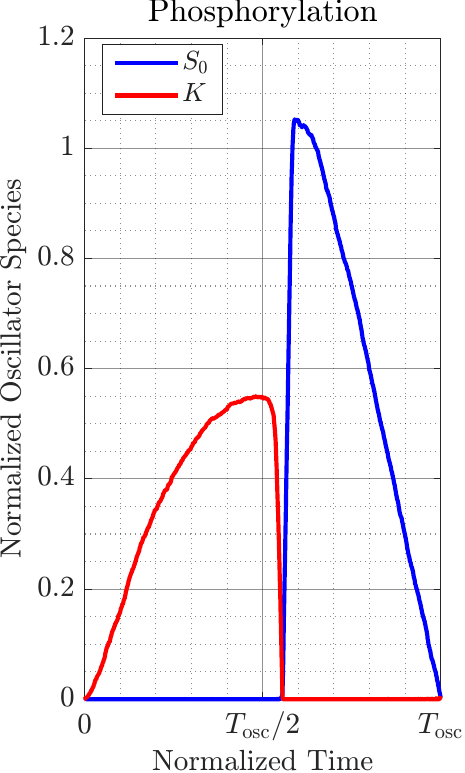}
    \caption{Example trajectories of three different oscillators over one normalized oscillation period.}
    \label{fig:osc_trajectories}
\end{figure}

\subsection{AM Cell-Cycle Switch Oscillator}

The first candidate is a coupled \textit{2-AM Switch} oscillator that is based on a cell-cycle-like structure and the common \ac{AM} algorithm~\cite{cardelliCellCycleSwitch2012}.
Cell division must be controlled and executed at regular intervals with high reliability. The regulatory mechanism in biological system acts as a switch based on two \acp{PFBL}, where an activator and inhibitor enzyme interact with each other~\cite{cardelliEfficientSwitchesBiology2017a}. The structure of the cell cycle switch is shown in Figure~\ref{fig:osc_diagrams}.
Expressing its fundamental behavior, the cell cycle switch can be reduced to the \ac{AM} decision algorithm from \ac{CRN} computing, which from two inputs converts all molecules to the species which was initially in the majority.
Coupling two \ac{AM} switches with \acp{NFBL} has been shown to produce sustained oscillations~\cite{cardelliCellCycleSwitch2012}. 
The model is parametrized mainly by an internal \ac{RRC} $\kappa_\mathrm{i}$ and an external coupling \ac{RRC} $\kappa_\mathrm{e}$ and oscillations are obtained for relatively small coupling compared to the internal speed~\cite{cardelliCellCycleSwitch2012}. In this case, we set $\frac{\kappa_\mathrm{e}}{\kappa_\mathrm{i}} = 0.2$. The full set of reactions and parameters used can be found in the supplementary material.
Figure~\ref{fig:osc_trajectories} depicts one example period of the 2-AM switch oscillator on the left side. The selected complementary species $X_1$ and $Y_1$ correspond to the activator and inhibitor species of one of the AM switches.

\subsection{Pentilator}

The second candidate is the gene-regulatory Pentilator, a five-gene variant of the basic Repressilator oscillation model. A gene-regulatory oscillator consists of a sequence of genes and proteins, where the protein product of each gene acts as a repressor for the next gene in the cycle forming an \ac{NFBL}~\cite{tsaiRobustTunableBiological2008}. The Repressilator was the first artificially constructed biological oscillating network within living cells~\cite{elowitzSyntheticOscillatoryNetwork2000}. 
The transcription process is most commonly modeled explicitly using Hill-type propensity functions, i.e. variable \acp{RRC}, where each gene transcription reaction has a reaction "speed" of the form
    \begin{equation}
        f = \frac{\alpha}{1 + P^n}, 
    \end{equation}
with the repressor protein $P$ being the preceding protein in the cycle, a scaling parameter $\alpha$, and the Hill coefficient $n$. In addition, the oscillator is parametrized by a production rate $\beta$ of the proteins from mRNA, as well as an mRNA degradation rate $d$, which can be used to modulate the frequency.
The five-gene Pentilator variant has been shown to exhibit reduced amplitude and frequency variability~\cite{tsaiRobustTunableBiological2008}, which is why we choose it here. A simplified structural diagram of the cycle from gene $G1$ to $G5$ can be seen in Figure~\ref{fig:osc_diagrams}.
For the evaluation, we set $\alpha = 1000$ and the Hill coefficient to $n=2$. The full model and parameters can be found in the supplementary material.
For the time-domain example period in Figure~\ref{fig:osc_trajectories}, two representative protein trajectories with minimal overlap are chosen.

\subsection{Mixed-Mechanism Phosphorylation}

As a third candidate we choose an enzyme-mediated dual-site phosphorylation network, which is derived from natural processes that involve switching complexes such as enzymes and proteins from active to inactive states~\cite{cardelliCellCycleSwitch2012, Mixed_mechs_of_multi-site_phosphor}.
Phosphorylation networks describe a process in which a substrate is modified by two counteracting enzymes: a kinase and a phosphatase~\cite{Mixed_mechs_of_multi-site_phosphor}. The kinase forms the complex by attaching phosphate groups to the substrate site (phosphorylation), while the phosphatase removes them (dephosphorylation). There are different mechanisms of phosphorylation depending on whether the substrate is added and removed individually, or first all are added and then removed. A process combining variants of both is called \textit{mixed-mechanism}.
Studies have shown that within specific molecular ratios some mixed-mechanism models are capable of generating oscillations~\cite{conradiEmergenceOscillationsMixedMechanism2019, Mixed_mechs_of_multi-site_phosphor}. A simplified diagram of the oscillator is shown in Figure~\ref{fig:osc_diagrams}.
We have selected such an oscillating model from~\cite{Mixed_mechs_of_multi-site_phosphor}, where a parameter configuration with a set of \acp{RRC} is suggested, which we have slightly modified for ease of uniform scaling. By scaling the \acp{RRC} uniformly the oscillation frequency is modulated. The full parameter set and model can be found in the supplementary material.
In Figure~\ref{fig:osc_trajectories}, we have chosen two species with minimum overlap, which represent the first substrate in the phosphorylation chain $S_0$, and the kinase $K$.

\subsection{Evaluation and Selection}
\begin{figure}
    \centering
    \includegraphics[width=0.48\linewidth]{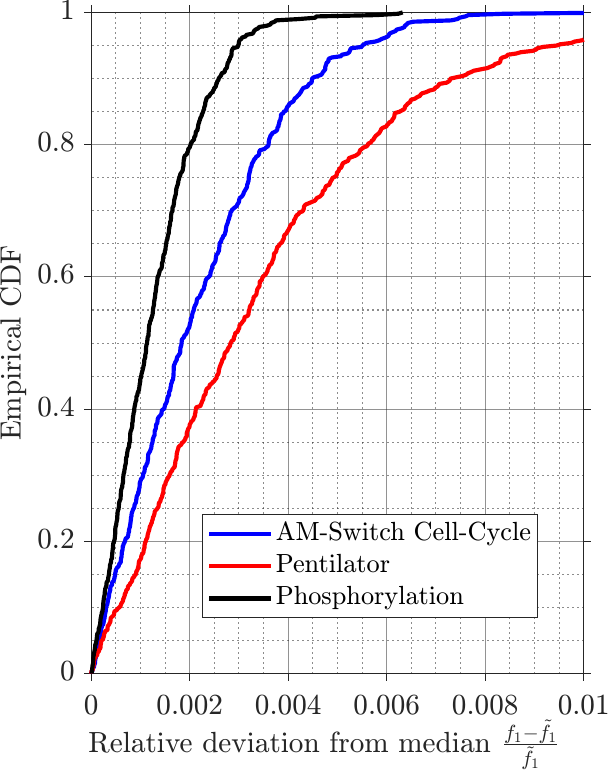}\hfill\includegraphics[width=0.48\linewidth]{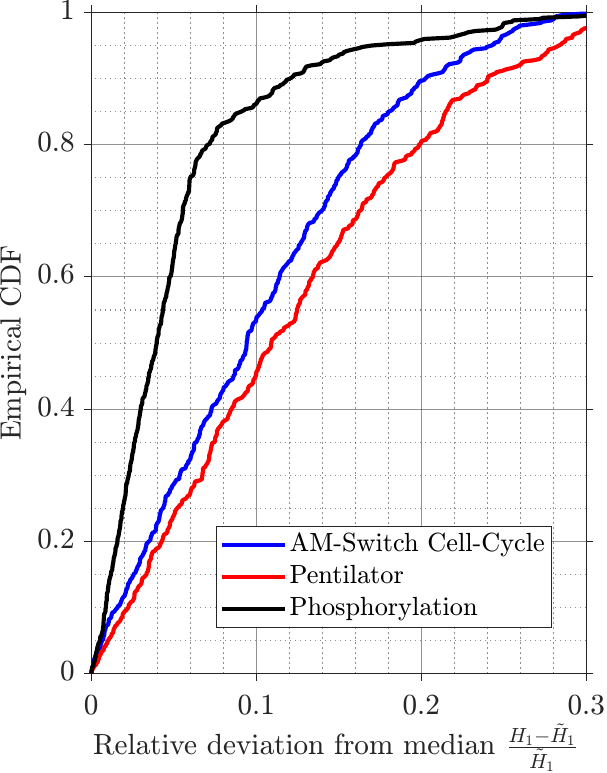}
    \caption{Empirical CDFs from stochastic simulations of the three considered oscillator models. CDF shows the fraction of trajectories, where the relative deviation from the median fundamental frequency $\tilde{f}_1$ and fundamental amplitude $\tilde{H}_1$ is lower than the $x$-axis value.}
    \label{fig:oscillator_cdfs}
\end{figure}

Firstly, in Figure~\ref{fig:osc_trajectories}, we can visually compare the representative trajectories of the three oscillators.
While all three exhibit the general desired pattern of oscillating from near-zero to the peak in two molecule species that are offset by half a cycle, there are some notable differences. For example, the Pentilator trajectories seem to showcase a significant amount of noise, while the phosphorylation oscillator appears the most stable. For the latter, we also see the least amount of inter-species overlap, while for the 2-AM and Pentilator variant, there is a significant part of the cycle where both species are present.
Secondly, we can take a look at the systematic analysis in Figure~\ref{fig:oscillator_cdfs}. Here, we extract the median values of the fundamental frequency and fundamental amplitude, $\tilde{f}_1$ and $\tilde{H}_1$, from the simulated trajectories and plot an empirical \ac{CDF} of the relative deviations $\frac{f_1 - \tilde{f_1}}{\tilde{f_1}}$, $\frac{H_1 - \tilde{H_1}}{\tilde{H_1}}$. In the plot, a line that is further to the top left represents higher stability. The results show  the phosphorylation oscillator with the highest stability for both $f_1$ and $H_1$ by a significant margin, while the Pentilator exhibits the largest deviations, and the 2-AM oscillator in the middle.
Given the two properties of smallest inter-species overlap and highest frequency and amplitude stability, we select the phosphorylation oscillator as the chemical clock we will use for ChemSICal-Net. We will use the two species shown in Figure~\ref{fig:osc_trajectories} to define activation windows for the different \ac{SIC} stages and evaluate the impact on performance as we will define it in Section~\ref{sec:simulation}.

\section{ChemSICal-Net CRN Design}\label{sec:crn_design}

\begin{figure*}
    \centering
    \includegraphics[width=\linewidth]{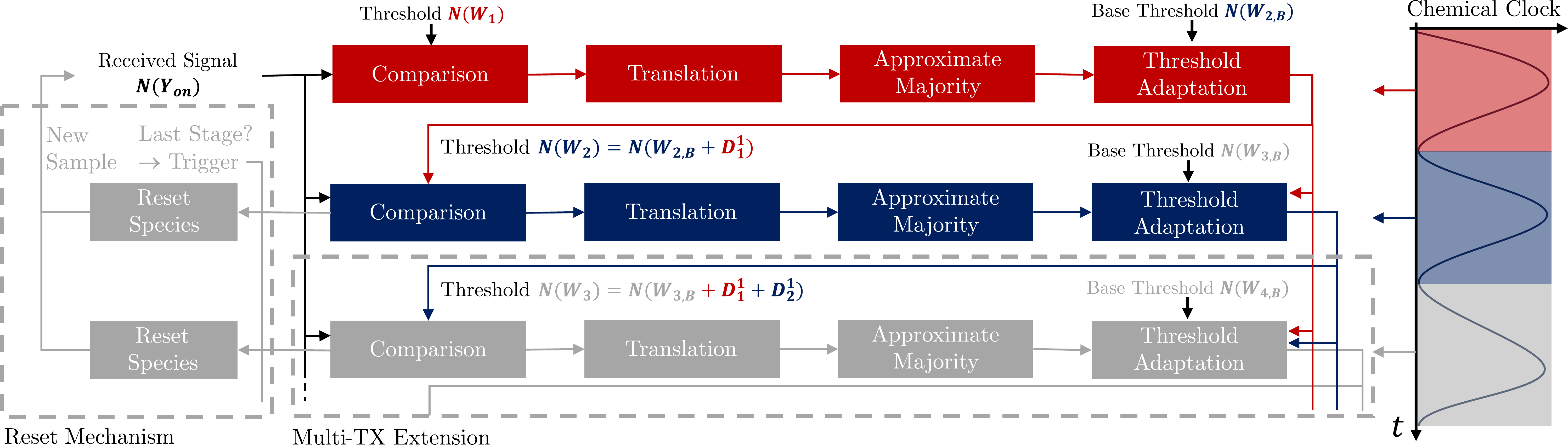}
    \caption{Proposed ChemSICal-Net block diagram: A \ac{CRN} implementing a multi-stage \ac{SIC} algorithm. Stages are controlled by a chemical oscillator. Optional extensions in grey include a reset mechanism and the addition of more \acp{TX}.}
    \label{fig:chemsical-net}
\end{figure*}

\begin{figure}
    \centering
    \includegraphics[width=0.45\linewidth]{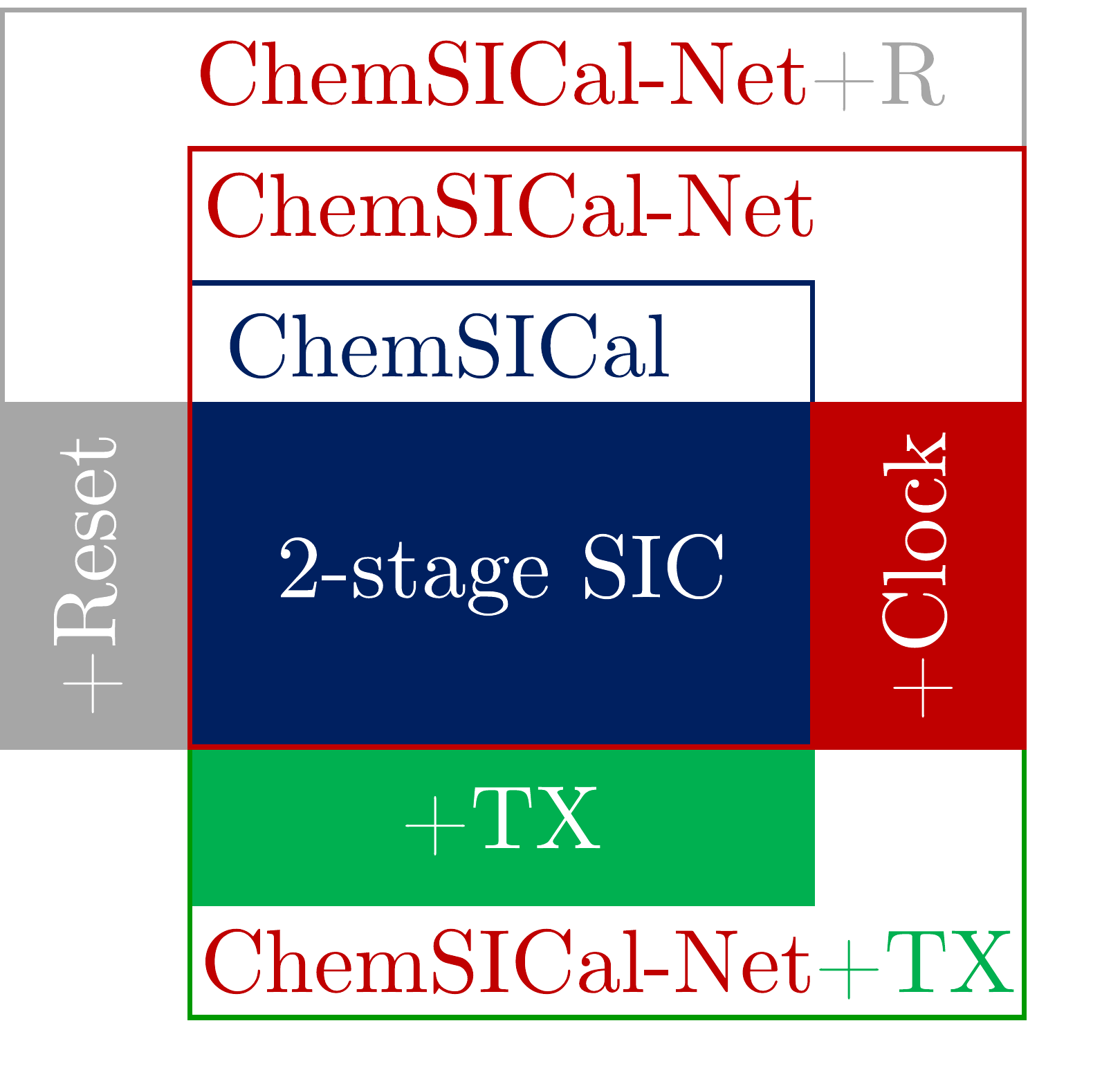}
    \caption{Compact overview, mirroring Figure~\ref{fig:chemsical-net}, of the different model variants utilized throughout the paper and their functional blocks.}
    \label{fig:model_overview}
\end{figure}

In this section, we will introduce the baseline ChemSICal design and its timed extension ChemSICal-Net, by introducing the chemical block structure underlying the \ac{SIC} computation, as well as the interface with the phosphorylation oscillator. Additionally, we describe a proof-of-concept reset mechanism and its integration into the system (ChemSICal-Net+R), and we describe how the structure can be conceptually scaled to many \acp{TX} and how the complexity of the system changes (ChemSICal-Net+TX). The full block diagram of the system design including all extensions can be found in Figure~\ref{fig:chemsical-net}. An overview of all the models, the different associated extensions, and their names can be seen in Figure~\ref{fig:model_overview}.

The objective of ChemSICal is to implement the \ac{SIC} detection logic using only chemical reactions. To that end, the observation $n_\mathrm{s}$ from the \ac{DBMC} channel is converted into a chemical representation and processed by a \ac{CRN}~\cite{wietfeldChemSICalEvaluatingStochastic2025}.
We represent the sampled input $n_\mathrm{s}$ as a discrete molecule count of a designated input species $Y_\mathrm{on}$, i.e. $N(Y_\mathrm{on}) = n_\mathrm{s}$, which can be interpreted as the number of signaling molecules that enter the reaction volume at the \ac{RX}. An implicit assumption is the existence of a sample-and-hold mechanism, which captures the number of molecules within the \ac{RX}. This could be addressed in future work, for example, by assuming more complex receptors such as ligand-binding types~\cite{farsad_comprehensive_2016}.
The chemical \ac{SIC} computation then produces two indicator species per TX$_i$ denoted by $D_i^0$ and $D_i^1$ in a comparison and decision mechanism which uses the input species $Y_\mathrm{on}$ and a \ac{TX}-specific threshold species $W_i$. For the bit detected via ChemSICal, we adopt the established decision rule
    \begin{equation}\label{eq:decision_rule}
        \hat{s}_i^\mathrm{C} = 
        \begin{cases}
            1 \qquad N(D_i^1) \geq N(D_i^0),\\
            0 \qquad \mathrm{otherwise,}
        \end{cases}
    \end{equation}
which directly mirrors the binary threshold decisions of the analytical \ac{SIC} from Section~\ref{subsec:noma_sic}.
The \ac{CRN} is parametrized by numerous additional intermediate species and a set of \acf{RRC}, describing the speed of each reaction. The performance of the chemical \ac{RX} depends critically on these reaction parameters rather than only on the analytically optimal threshold parameters~\cite{wietfeldChemSICalEvaluatingStochastic2025}.

The baseline two-stage \ac{SIC} logic is decomposed into modular reaction blocks that implement comparison, translation, \acf{AM}, and threshold adaptation. Each \ac{TX} corresponds to one stage.
The second-stage blocks reuse the same constant $N(Y_\mathrm{on})$, as it is not consumed, and adapt the threshold for TX$_2$, $N(W_2)$, according to the detection result in stage 1. We make use of the following basic \ac{CRN} blocks, as defined in~\cite{vasicCRNMolecularProgramming2020}:

\textit{Comparison:} Compares an input species ($Y_\mathrm{on}$) to a threshold ($W_i$) and generates two indicator molecule species ($X_{\mathrm{on},i}$, $X_{\mathrm{off},i}$), the ratio of which indicates the ratio between input and threshold.

\textit{Translation:} Generates detection species ($D_i^0$, $D_i^1$) of the exact same number of molecules as one or multiple inputs ($X_{\mathrm{on},i}$, $X_{\mathrm{off},i}$). This is used to decouple subsequent reactions, i.e., limit interference of the previous reaction on the reactants of the following ones.

\textit{Approximate Majority (\ac{AM}):} Takes two inputs ($D_i^0$, $D_i^1$) and amplifies the species that was initially in the majority, i.e., turns $N(D_i^j) > N(D_i^k)$ into outputs $N(D_i^j) \gg N(D_i^k)$. As a results, many more of one species, and almost none of the other are left, creating an approximate binary decision.

\textit{Threshold Adaptation:} Corresponds to an addition \ac{CRN}, similar to Figure~\ref{fig:crn_example}. The base threshold species molecule count, $N(W_{2,\mathrm{B}})$, and the detection species molecule count $N(D_1^1)$ are added to form the threshold species $N(W_2)$. This links the first stage for detecting the bit from TX$_1$ with the second stage for the bit from TX$_2$. The molecule counts of each species must be tuned such that: $N(W_{2,\mathrm{B}}) = \tau_2^0$ and $N(X_{\mathrm{on},1}) + N(X_{\mathrm{off},1}) + N(W_{2,\mathrm{B}}) = \tau_2^1$. The latter must hold, since the asymptotic result of the first-stage \textit{\ac{AM}} block in case of $\hat{s}_1^C = 1$ is $N(D_1^1) \approx N(X_{\mathrm{on},1}) + N(X_{\mathrm{off},1})$.

The full set of chemical reactions per block and for the reset and third \ac{TX} extension introduced later can be found in the supplementary material.

\begin{figure}
    \centering
    \includegraphics[width=0.8\linewidth]{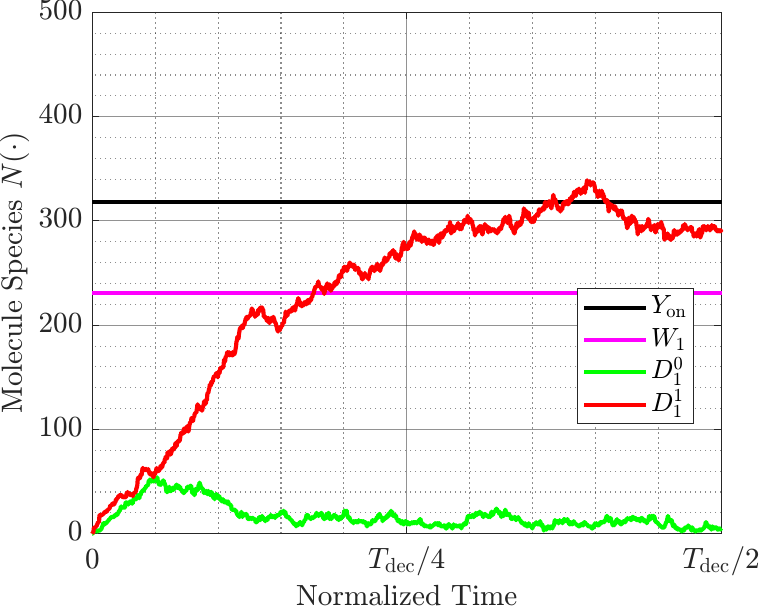}
    \caption{Example trajectory of the molecule species relevant to the first stage for input $N(Y_\mathrm{on}) = 318$, showing the decision variables $D_1^0, D_1^1$ converging according to the relation between input and threshold (here, $N(Y_\mathrm{on})>N(W_1)$).}
    \label{fig:trajectory_single_stage}
\end{figure}

\begin{figure}
    \centering
    \includegraphics[width=0.8\linewidth]{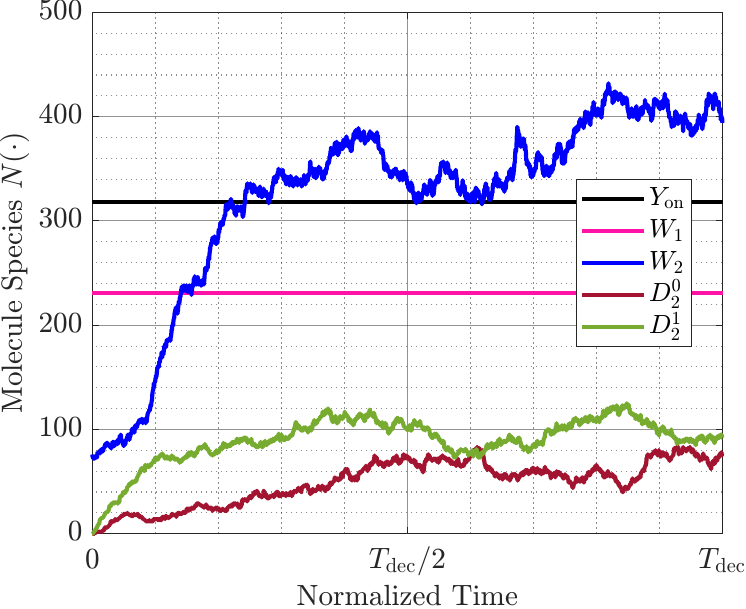}
    \caption{Example trajectory of the molecule species relevant for the second stage for input $N(Y_\mathrm{on}) = 318$, for the configuration \textit{without} timing control, i.e. all stages are running simultaneously. The plot shows the molecule counts of $D_2^0, D_2^1$ converging \textit{incorrectly} since the threshold $N(W_2)$ needs to be adapted in reaction to the first-stage detection, but the second stage already starts computing based on the wrong value of $N(W_2)$.}
    \label{fig:trajectory_always_on}
\end{figure}

Each reaction block is parametrized by one \ac{RRC} that defines the speed of execution of that block. According to each block, we define the \acp{RRC} per TX$_i$/stage $i$: $\kappa_{\mathrm{C},i}$ for \textit{Comparison}, $\kappa_{\mathrm{Tr},i}$ for \textit{Translation}, $\kappa_{\mathrm{AM},i}$ for \textit{\ac{AM}}, and $\kappa_{\mathrm{TA},i}$ for \textit{Threshold Adaptation}.
We maintain the single-volume assumption for the entire \ac{CRN}, meaning there is no need for separate compartments, but a requirement for correct temporal ordering via chemical design.
In the baseline ChemSICal model~\cite{wietfeldChemSICalEvaluatingStochastic2025}, the ordering must be implicitly achieved by tuning the \acp{RRC} across stages so that stage-2 computations take place after stage-1 computations have finished.
Figure~\ref{fig:trajectory_single_stage} shows an example trajectory for the input $Y_\mathrm{on} = 318$, we can see how the species for Y and W1 remain constant and the decision variables react in response to the input, showing $D_1^1$ as the clearly larger concentration at the end.
In Figure~\ref{fig:trajectory_always_on}, we show the entire \ac{SIC} cycle until the decision time $T_\mathrm{dec}$, for the case of the baseline ChemSICal where both stages run simultaneously. The parameters for this case are optimized using the scheme later presented in Section~\ref{sec:optimization}. We can see that the second symbol is actually incorrectly detected. The threshold $N(W_2)$ is adapted, i.e. increased, since we detect a '1' for TX$_1$. Initially $N(Y_\mathrm{on})>N(W_2)$, and the second stage starts to produce more $D_2^1$ molecules (see dark green line). However, the adaptation causes $N(W_2)$ to exceed $N(Y_\mathrm{on})$ eventually, which means $D_2^0$ should dominate the detection. However, the second stage ends up in an unstable and undecided state, where $N(D_2^0)\approx N(D_2^1)$. The main reason lies in the fact that the second stage already reacts to the incorrect thresholds at the very beginning and does not correct itself later, when the threshold has changed. In the next section, we will discuss the extension to explicit ordering using a chemical clock, which can help in this case.

\subsection{Chemical Clock Integration - ChemSICal-Net}

The novel model variant ChemSICal-Net (see Figure~\ref{fig:model_overview}) integrates a chemical oscillator that generates a repeatable phase signal.
This phase signal is mapped to \ac{SIC} stages such that reactions are predominantly active in intended time windows, rather than competing immediately from $t=0$.

As discussed in Section~\ref{sec:oscillators}, we select the dual-site phosphorylation oscillator, which provides multiple oscillating species with minimal inter-species overlap, e.g., $S_0$ and the kinase $K$.
To gate stage~2 (blue row in Figure~\ref{fig:chemsical-net}), we multiply the propensity function of each reaction block in that stage by a common Hill activation factor based on $S_0$,
\begin{equation}
    g_{S_0}(t) = \frac{N(S_0)(t)^{n_{S_0}}}{N(S_0)(t)^{n_{S_0}} + S_{0,\mathrm{half}}^{n_{S_0}}},
\end{equation}
with Hill coefficient $n_{S_0}$ and half-activation value $S_{0,\mathrm{half}}$, which jointly determine the steepness of the clock-induced activation transition. Similarly, we can define a Hill activation factor $g_K(t)$ based on the kinase $K$.

\begin{figure}
    \centering
    \includegraphics[width=0.8\linewidth]{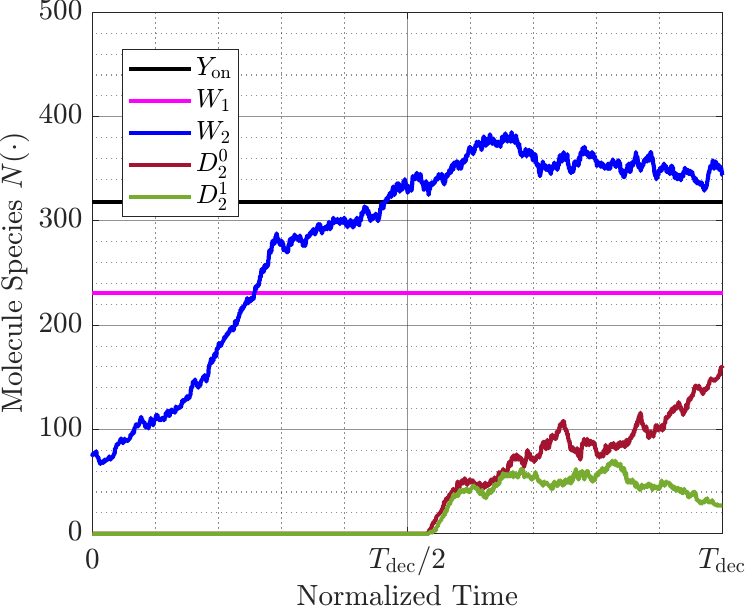}
    \caption{Example trajectory of the molecule species relevant for the second stage for input $N(Y_\mathrm{on}) = 318$, for the configuration \textit{with} timing control. The plot shows the decision variable molecule counts of $D_2^0, D_2^1$ are converging \textit{correctly} (in contrast to Figure~\ref{fig:trajectory_always_on}), since the threshold $N(W_2)$ can be adapted in the first half of the decision time, and the second-stage reactions start computing only in the second half based on correct input values.}
    \label{fig:timing_control}
\end{figure}

We illustrate the timing effect in Figure~\ref{fig:timing_control} using the same input and species selection as in Figure~\ref{fig:trajectory_always_on}.
Here, we define the \emph{decision horizon} $T_{\mathrm{dec}}$ as the time at which the \ac{CRN} output is read out (see Section~\ref{subsec:metrics}).
For timed variants, we tune the oscillator such that one full period equals the decision horizon, i.e., $T_{\mathrm{osc}} = T_{\mathrm{dec}}$.
Consequently, for the two-stage case, stage~2 is activated predominantly during the second half-cycle, starting at $T_{\mathrm{osc}}/2 = T_{\mathrm{dec}}/2$.
The shown trajectory yields a correct final decision and highlights how the threshold $W_2$ is first adapted based on detection stage~1 (see Figure~\ref{fig:trajectory_single_stage}) before the second stage is enabled.

Overall, ChemSICal-Net provides a tunable decision time per symbol by adjusting $T_{\mathrm{dec}}$ (and thus $T_{\mathrm{osc}}$ in timed variants):
smaller $T_{\mathrm{dec}}$ increases throughput but risks incomplete reaction processing, whereas larger $T_{\mathrm{dec}}$ improves \ac{CRN} reliability at the cost of latency.

\subsection{Further Extensions}

\subsubsection{Additional TXs}\label{subsubsec:crn_design_additional_txs}

\begin{figure}
    \centering
    \includegraphics[width=0.85\linewidth]{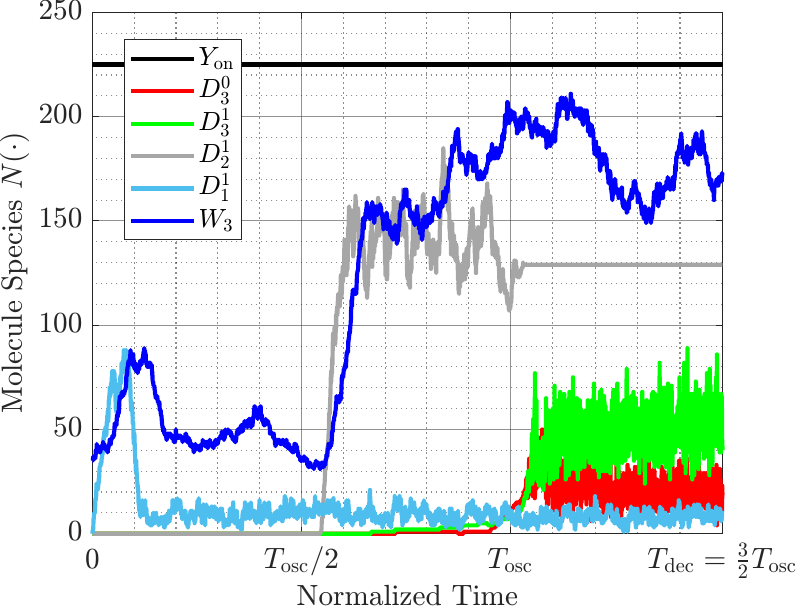}
    \caption{Example trajectory of the molecule species relevant for the third stage for an input of $N(Y_\mathrm{on})=225$. The plot shows the threshold $N(W_3)$ is adapted based on the decision species for stage 1 (light blue) and stage 2 (grey) and ultimately leads to the correct decision for stage 3, since $N(Y_\mathrm{on})>N(W_3)$.}
    \label{fig:3tx_trajectory}
\end{figure}

The ChemSICal-Net block structure can be extended beyond two \acp{TX} by repeating the stage pattern consisting of \textit{Comparison}, \textit{Translation}, and \textit{\ac{AM}} for each additional \ac{TX} and by introducing an additional \textit{Threshold Adaptation} step per added stage.
For a third \ac{TX} (TX$_3$), we introduce a third stage with threshold species $W_3$ and decision species $D_3^0$ and $D_3^1$, and the corresponding computation blocks are indicated by the shaded \textit{Multi-TX Extension} region in Figure~\ref{fig:chemsical-net}.
The key additional operation is the formation of the stage-3 threshold $W_3$ from a base threshold $W_{3,\mathrm{B}}$ and the detection evidence produced by the prior stages, where the conceptual structure in Figure~\ref{fig:chemsical-net} corresponds to an additive update of the form $N(W_3) = N(W_{3,\mathrm{B}}) + N(D_1^1) + N(D_2^1)$.
The molecule counts of $W_{3,\mathrm{B}}$ and the effective contribution of the decision evidence are tuned such that the resulting chemical thresholds match the intended analytical \ac{SIC} thresholds along the relevant decision branches.
To prevent premature computation with a non-adapted threshold, the stage-3 reaction blocks are activated in a dedicated time window after stage 1 and stage 2 have completed their threshold adaptation and decision formation.
In the example structure in Figure~\ref{fig:chemsical-net}, the stage-2 and stage-3 blocks are controlled by different clock phases so that the three stages are executed sequentially, each within one oscillation period without overlap.

In terms of complexity growth, for $M$ \acp{TX}, ChemSICal-Net comprises $M$ copies of the \textit{Comparison, Translation, \ac{AM}} blocks plus $M-1$ \textit{Threshold Adaptation} blocks, i.e., $B(M)=3M+(M-1)=4M-1$ logical blocks and thus reactions/species scale linearly with $M$.
However, for threshold formation, stage $i$ aggregates the base threshold and the prior decision evidence,
\begin{equation}
N(W_i)=N(W_{i,\mathrm{B}})+\sum_{j=1}^{i-1}\,N(D_j^1),
\end{equation}
so adding one \ac{TX} increases the threshold adaptation by one additional evidence term per subsequent stage, which leads to polynomial growth. The exponential growth that is usually associated with a binary tree decision mechanism (see \ac{SIC} structure in Figure~\ref{fig:sic}) can be avoided, due to the design of ChemSICal-Net as a successive aggregator of bit decisions for threshold adaptation.
We treat the 3-\ac{TX} variant as a proof-of-concept extension to demonstrate feasibility and scaling trends, and we restrict the detailed optimization study to a representative configuration.
Figure~\ref{fig:3tx_trajectory} illustrates an example trajectory excerpt, highlighting two consecutive threshold-adaptation steps and the delayed activation of the third-stage decision dynamics.

\subsubsection{Reset Mechanism}\label{subsubsec:crn_design_reset}

\begin{figure}
    \centering
    \includegraphics[width=0.8\linewidth]{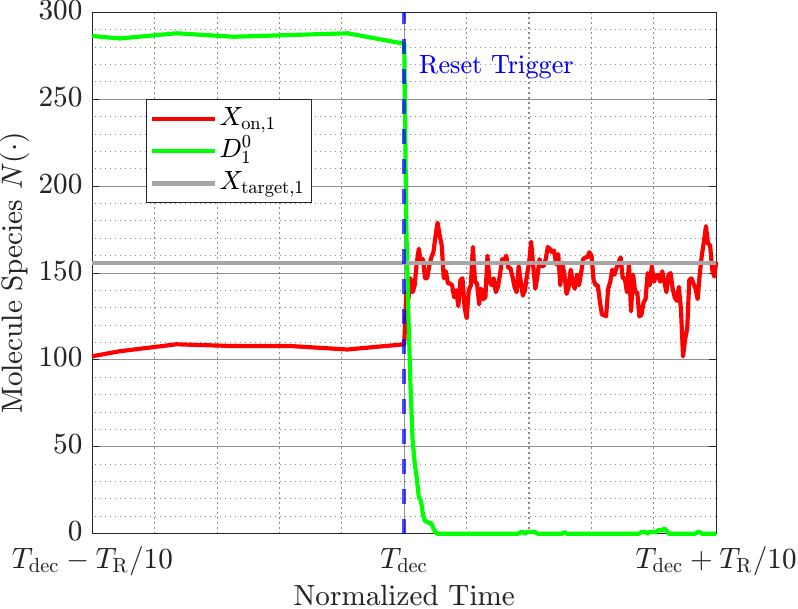}
    \caption{Example trajectory of the molecule species relevant for the reset mechanism. We show a small snippet just before and just after the reset trigger. We show a selected decision species $D_1^0$ returning to a molecule count of zero, and the intermediate comparison species molecule count $N(X_\mathrm{on,1})$ being reset to its target value within a short amount of time.}
    \label{fig:reset_trajectory}
\end{figure}

The baseline ChemSICal and ChemSICal-Net models are one-shot computations, because internal intermediates and the \textit{\ac{AM}} outputs accumulate over time and settle into strongly imbalanced states that cannot be reused without reinitialization.
ChemSICal-Net+R introduces an explicit reset phase that restores selected internal species to well-defined initial conditions at the end of each symbol period, enabling repeated symbol processing with the same reaction volume.
We distinguish two reset actions, namely clearing species that should return to zero (e.g., decision and intermediate species) and restoring species that should return to fixed reference values.
The reset is controlled by a dedicated trigger species that activates reset reactions only during a designated reset window, as indicated by the dashed \textit{Reset Mechanism} block in Figure~\ref{fig:chemsical-net}. This trigger is not explicitly modeled for this proof-of-concept evaluation but it could be connected to an appropriate oscillation species similar to the chemical clock with a slower frequency to activate only after all stages have been computed.
For clearing, the trigger activates reactions that convert targeted species into inert waste products, which effectively removes residual molecules without altering the held input $Y_\mathrm{on}$ during the decision phase.
For restoration to reference values, the trigger activates reactions that release reference molecules from protected reservoir species into the active pool, thereby recreating the desired initial molecule counts without requiring external injection of each individual species.
In the timed design, the reset window is scheduled after all detection stages have completed, so that reset reactions do not interfere with threshold adaptation or decision formation.
Figure~\ref{fig:reset_trajectory} illustrates an excerpt of an example reset cycle in which, $D_0^1$, a decision-related species is driven back to zero and, $X_\mathrm{on,1}$, a reference species is restored to its target molecule count before the subsequent computation begins. The overall reset cycle lasts for time $T_\mathrm{R}$.

\section{Simulation Setup}\label{sec:simulation}

\begin{table*}[t]
\caption{Communication System Parameters}
\label{tab:sys_params_main}
\centering\scriptsize
\begin{tabular}{lll}
\hline
\textbf{Parameter}    & \textbf{Symbol}                            & \textbf{Values}                          \\ \hline
Number of TXs            & $M$                                        & $2, 3$                                     \\
TX distances          & $\{d_1, d_2, d_3\}$                          & $\{10,12,14\}\,\unit{\micro\meter}$                \\
RX radius             & $r$                                        & $\qty{1}{\micro\meter}$                 \\
Diffusion coefficient & $D$                                        & $\qty{1e-9}{\meter\squared\per\second}$ \\
Molecules sent per bit-1                                & $N_\mathrm{TX}$ & $\qty{1e6}{\mathrm{molecules}}$                                  \\ 
\ac{SIC} thresholds ($K=2$)                      & $\{\tau_1, \{\tau_2^0, \tau_2^1\}\}$  & \{231, \{78, 386\}\}\,$\unit{\mathrm{molecules}}$           \\
\ac{SIC} thresholds ($K=3$)                      & $\{\tau_1, \{\tau_2^0, \tau_2^1\}, \{\tau_3^{00}, \tau_3^{01}, \tau_3^{10}, \tau_3^{11}\}\}$  & \{267, \{114, 422\}, \{35, 192, 343, 500\}\}\,$\unit{\mathrm{molecules}}$           \\
\hline
\end{tabular}%
\end{table*}

In this section, we define the performance metrics used to evaluate our system, describe the solvers and \acp{SSA} used to simulate the \ac{CRN}, and outline the \ac{MCS} evaluation structure.

\subsection{Solvers and Simulation Methods}\label{subsec:solvers}

The evaluation metrics are based on the chemical detection outcomes $\hat{s}_i^\mathrm{C}$ and therefore on the terminal molecule counts of the detection species pairs $(D_i^0,D_i^1)$ as defined in the system model.
The correctness of the \ac{CRN} outcome depends on the input value $N(Y_{\mathrm{on}})$, i.e., on the sampled received signal $n_{\mathrm{s}}$, which dictates the expected outcome according to the non-chemical \ac{SIC} decision rule.
We evaluate the \ac{CRN} using both a deterministic solver and a stochastic solver, because the deterministic mean-field behavior alone does not capture the probabilistic dynamics that emerge at finite molecule counts.
The deterministic solver converts the \ac{CRN} into a set of \acp{ODE} and computes continuous time-varying concentration trajectories for all species, which is useful for qualitative validation and debugging of reaction ordering.
The stochastic solver applies an \ac{SSA} based on Gillespie's algorithm, which generates statistically accurate sample trajectories of the discrete-state \ac{CRN} dynamics.
In each \ac{SSA} trajectory, intrinsic molecular noise is reflected by random reaction event times driven by stochastic molecule interactions, and repeated trajectories are therefore required to characterize detection performance statistically.
Additive external molecular noise sources such as leakage molecules are not modeled in the present setup, and their inclusion is expected to primarily shift effective thresholds and error regions rather than alter the qualitative operation.

\subsection{Performance Metrics}\label{subsec:metrics}

For a given input molecule count $N(Y_{\mathrm{on}})$, we define a binary correct-detection indicator as $c[N(Y_{\mathrm{on}})]\in\{0,1\}$ for deterministic evaluation and as $c_j[N(Y_{\mathrm{on}})]\in\{0,1\}$ for a single \ac{SSA} trajectory $j$.
A trajectory is considered correct if the \ac{CRN} returns the correct joint decision $[\hat{s}_1^C\ \hat{s}_2^C]$ for the given input $N(Y_{\mathrm{on}})$, where each $\hat{s}_i^\mathrm{C}$ is obtained by comparing the terminal counts of $(D_i^0,D_i^1)$ according to the decision rule (\ref{eq:decision_rule}) in the system model.
For reset-enabled variants, the per-trajectory correctness check additionally requires that designated species are returned to their intended post-run target state within a tolerance band, so that correct decisions are only counted if the run is both correct and reusable.
The probability of correct detection for a given input and $N_\mathrm{traj}$ different trajectories $j$ is estimated as
    \begin{equation}
        P_{\mathrm{d}}[N(Y_{\mathrm{on}})] = \frac{1}{N_{\mathrm{traj}}}\sum_{j=1}^{N_{\mathrm{traj}}} c_j[N(Y_{\mathrm{on}})].
    \end{equation}
The primary aggregate metric is an input-weighted probability of error, where errors are weighted by the input \ac{PMF} $p_{n_{\mathrm{s}}}[N(Y_{\mathrm{on}})]$ such that unlikely inputs have less influence than likely ones.
We compute the input-weighted error probability as
    \begin{equation}
        P_{\mathrm{e}} = 1 - \sum_{N(Y_{\mathrm{on}})} p_{n_{\mathrm{s}}}[N(Y_{\mathrm{on}})]\cdot P_{\mathrm{d}}[N(Y_{\mathrm{on}})],
    \end{equation}
which is equivalent to weighting the correct-detection probability and converting to an error metric.
We additionally define a reduced input-weighted error probability by restricting the sum to a selected subset of input values that capture the high-likelihood region of the input distribution, which yields a more efficient objective for parameter search.
We define the decision time $T_\mathrm{dec}$, i.e., the time at which \ac{CRN} states are read out and mapped to $[\hat{s}_1^C\ \hat{s}_2^C]$, and $T_\mathrm{dec}$ is treated as a key design knob in the evaluation.

\begin{table}[t]
\centering
\caption{\ac{CRN} Initial Values}
\label{tab:crn_init_main}
\resizebox{\columnwidth}{!}{%
\begin{tabular}{lll}
\hline
Parameter          & Symbol          & Initial Value (before \ac{BO}) \\ \hline
Input Species      & $Y_\mathrm{on}$        & $0\leq N(Y_\mathrm{on}) \leq 600$ $\qty{}{\mathrm{molecules}}$                 \\
Comparison Species     & $\{X_{on,1}, X_{off,1}\}$      & 154, 154  $\qty{}{\mathrm{molecules}}$                 \\
                   & $\{X_{on,2}, X_{off,2}\}$      & 83, 84  $\qty{}{\mathrm{molecules}}$                  \\
                   & $\{X_{on,3}, X_{off,3}\}$     & 35, 36  $\qty{}{\mathrm{molecules}}$                  \\
Detection Species  & $\{D_i^0, D_i^1\}, i\in\{1,2,3\}$       & \{0,0,0,0\}  $\qty{}{\mathrm{molecules}}$                   \\
Threshold Species  & $W_{1}$         & 231  $\qty{}{\mathrm{molecules}}$                 \\
                   & $\{W_{2, B}, W_{2}\}$         & 78, 78  $\qty{}{\mathrm{molecules}}$                 \\
                   & $\{W_{3,B}, W_3\}$          & 35, 35  $\qty{}{\mathrm{molecules}}$                  \\
Helper Species     & $\{B_{1}, B_2, B_3\}$         & \{0,0,0\}  $\qty{}{\mathrm{molecules}}$                   \\                   
\hline

\end{tabular}%
}
\end{table}

\begin{table}[t]
\centering
\caption{\ac{CRN} Kinetic Parameters}
\label{tab:crn_rates_main}
\resizebox{\columnwidth}{!}{%
\begin{tabular}{lll}
\hline
Parameter          & Symbol          & Value \\ \hline
ChemSICal-Net \acp{RRC}     & $\kappa_{\mathrm{C},i}, \kappa_{\mathrm{Tr},i}, \kappa_{\mathrm{AM},i},$ & Before BO: see Table~\ref{tab:ODE_reaction_rates}\\
                   & $\kappa_{\mathrm{TA},i},\,i\in\{1,2,3\}$       & After BO: see Suppl. Mat.\\
Rate bounds & $\{\kappa_{\min},\,\kappa_{\max}\}$ & $\{10^{-3},\, 1\}$                  \\
Oscillator rates & $\kappa_1,\dots,\kappa_{10}$ & see Suppl. Mat.                   \\
Time reference & $T_\mathrm{ref}$ & 73 [unit-less]\\
Oscillation period & $T_\mathrm{osc}$ & $\{T_\mathrm{ref}/4, T_\mathrm{ref}/2, T_\mathrm{ref}, 2T_\mathrm{ref}\}$\\
Reset rates & $\kappa_{\mathrm{clear}},\kappa_{\mathrm{copy}},\kappa_{R,\mathrm{decay}}$ & see Suppl. Mat.               \\
Hill half-activation $K$ & $K_{\mathrm{half}}$ & $600$ $\qty{}{\mathrm{molecules}}$                \\
Hill half-activation $S_0$    & $S_{0,\mathrm{half}}$ & $600$  $\qty{}{\mathrm{molecules}}$               \\
Hill coefficients & $n_{K},\,n_{S_0}$ & $\{1,\,1\}$                  \\                
\hline
\end{tabular}%
}
\end{table}

\subsection{Monte-Carlo Simulations}\label{subsec:mc}

For each parameter configuration and each evaluated input molecule count $N(Y_{\mathrm{on}})$, we generate $N_{\mathrm{traj}}$ independent \ac{SSA} trajectories up to the decision time $T_\mathrm{dec}$ and compute $P_{\mathrm{d}}[N(Y_{\mathrm{on}})]$ as an empirical mean of correctness outcomes.
The overall input-weighted metric is obtained by combining the per-input estimates using the weights given by the input \ac{PMF} from the communication system model.
We report a 95\% \ac{CI} for Monte-Carlo estimates using a normal approximation for proportions with a $z$-multiplier of $1.96$, and we use the corresponding \ac{CI} half-width as an indicator of estimation reliability.
Unless stated otherwise, full evaluations use a high trajectory count per input to obtain stable $P_{\mathrm{d}}$ curves and reliable weighted error estimates, whereas optimization loops may use smaller budgets and increase $N_{\mathrm{traj}}$ only for promising or uncertain configurations.

\section{Optimization Scheme Comparison and Selection}\label{sec:optimization}

We treat the ChemSICal-Net system as a black-box optimization problem over a bounded parameter space because the objective is stochastic, non-convex, and only available as an estimate obtained from multiple \ac{SSA} runs.
Our goal is to minimize the likelihood-weighted error probability \(P_\mathrm{e}\), where correctness is estimated per input \(Y_\mathrm{on}\) from Monte-Carlo trajectories and then weighted by the input likelihood as defined in Section~\ref{sec:system_model}.
Equivalently, we maximize the detection accuracy \(1-P_\mathrm{e}\).

The optimization variables consist of the set of \acp{RRC} of the reaction blocks, denoted by \(\mathcal{K}\), as well as the initial molecule species concentrations, denoted by \(\mathcal{N}\).
This yields the constrained optimization problem
\begin{equation}
\begin{aligned}
& \underset{\mathcal{K},\,\mathcal{N}}{\text{minimize}}
& & P_\mathrm{e}(\mathcal{K}, \mathcal{N}) \\
& \text{subject to}
& & \kappa_\mathrm{min} \leq \kappa_i \leq \kappa_\mathrm{max}, \quad \kappa_i \in \mathcal{K},\\
& & & N_j \in \mathbb{Z}_{\ge 0}, \quad N_j \in \mathcal{N}.
\end{aligned}
\end{equation}
To solve this problem, we propose an adaptive \ac{BO} scheme as the default optimizer.
To evaluate its effectiveness against other state-of-the-art black-box methods, we additionally consider \ac{SA} and a \ac{MH}-\ac{MCMC} scheme, and we compare all methods under matched evaluation settings in a small set of representative scenarios.

\begin{figure}
    \centering
    \includegraphics[width=0.9\linewidth]{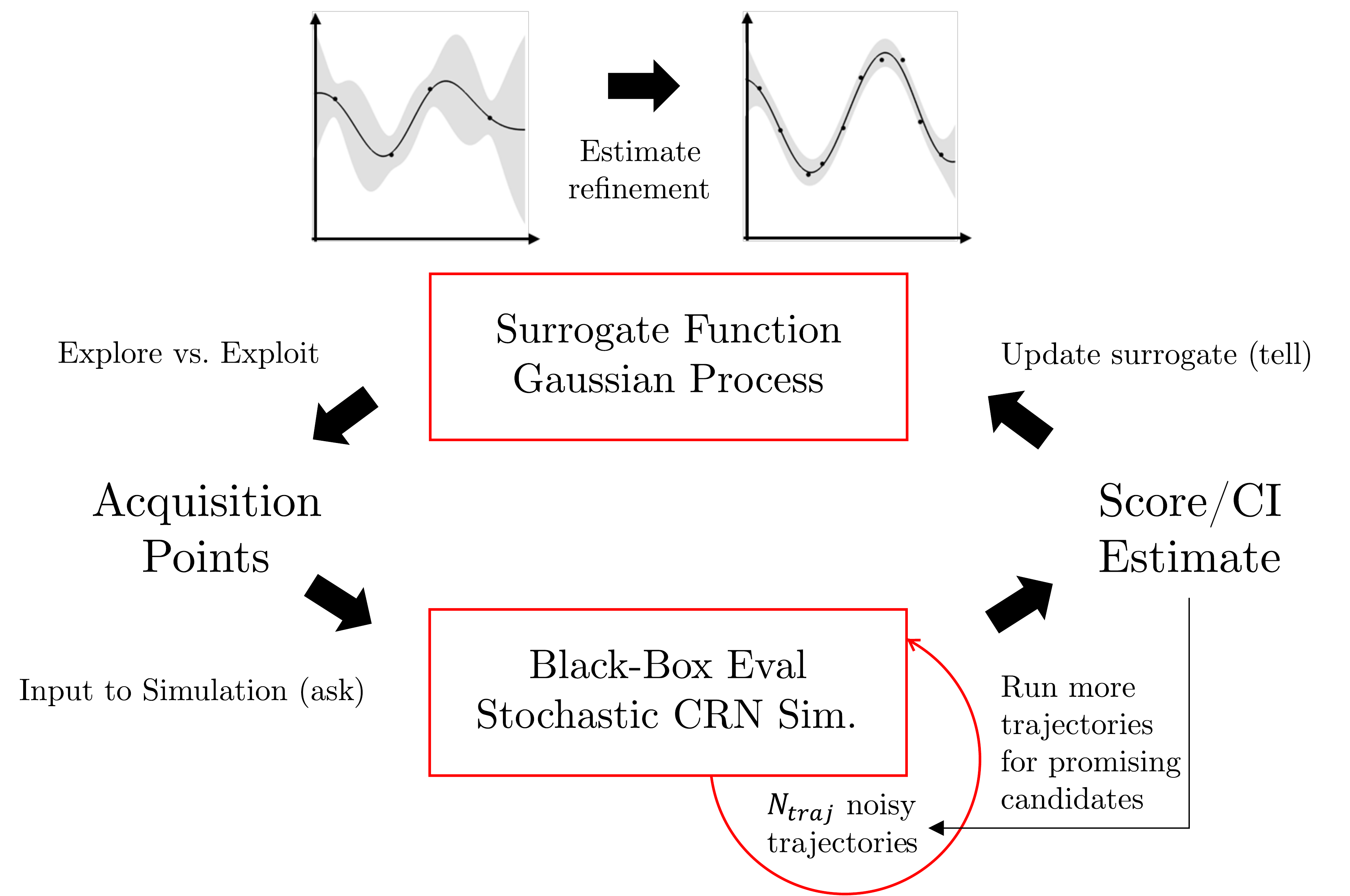}
    \caption{Illustration of the \ac{BO} loop, where noisy black-box evaluations are used to update a surrogate model, and an acquisition rule balances exploration and exploitation to select the next candidates.}
    \label{fig:bo_diagram}
\end{figure}

\subsection{Baseline Optimization Setup}\label{subsec:opt_baseline}

To reduce computational effort, we do not optimize over a dense sweep of possible \(Y_\mathrm{on}\) values.
Instead, we evaluate candidates on a sub-sampled input vector that is chosen to cover both (i) regions with high input likelihood and (ii) regions close to the relevant decision thresholds, where errors are most likely to occur.
This sub-sampling reduces the number of inputs per objective evaluation significantly, while still capturing the dominant contribution to the likelihood-weighted objective.

Since the objective is noisy due to finite Monte-Carlo sampling, each candidate must be evaluated using multiple \ac{SSA} trajectories.
To improve sample efficiency, we apply an adaptive replication scheme in which candidates are evaluated in successive rungs with increasing numbers of trajectories per input.
In our setup, we use the rung schedule \([20, 60, 100, 500]\) trajectories per input, such that the first rung enables fast screening and later rungs provide high-confidence evaluation for candidates that appear competitive.
A candidate is promoted to the next rung if it either reaches a rung-specific performance threshold (\(1-P_\mathrm{e} \geq 0.90\), \(0.95\), and \(0.99\) for successive rungs) or if its estimated uncertainty remains large (\ac{CI} half-width of $1-P_\mathrm{e}\ >0.1$), which prevents discarding candidates that might be strong but insufficiently sampled.

To enable a fair complexity comparison across optimization schemes, we report optimization performance against the cumulative number of simulated \ac{SSA} trajectories required to reach a given score, rather than against iteration count alone.
This is particularly important because adaptive replication makes the evaluation cost candidate-dependent.

\subsection{Adaptive Bayesian Optimization Scheme}\label{subsec:opt_bo}

Our \ac{BO} scheme follows an ask-tell loop with a Gaussian process surrogate model and an expected-improvement (EI) acquisition function.
The surrogate provides both a predicted objective value and an uncertainty estimate for each point in the parameter space, and the EI acquisition balances exploration of uncertain regions with exploitation of regions that are predicted to perform well. We use a Matérn kernel for the surrogate.

We evaluate candidates in batches to leverage parallel computation, while feeding all completed evaluations back into the surrogate model.
The resulting structure is summarized in Figure~\ref{fig:bo_diagram}, where the black-box objective evaluations update the surrogate, and the acquisition mechanism determines the next candidate parameters.

\subsection{Comparison with Benchmark Schemes}\label{subsec:opt_benchmarks}

\subsubsection{Simulated Annealing}\label{subsubsec:opt_sa}

We implement simulated annealing~\cite{kalivasAdaptionSimulatedAnnealing1995} as a multi-chain scheme to exploit parallelism and reduce sensitivity to initialization.
The method uses a temperature schedule that cools over the number of batch steps, enabling broader exploration early in the run and more local refinement later. We replicate the settings employed for \ac{CRN} optimization using \ac{SA} in related work~\cite{ashyraliyevSystemsBiologyParameter2009}.
Candidate proposals are generated as Gaussian steps in \(\log_{10}\) space for log-scaled rate parameters, which corresponds to multiplicative perturbations in the original parameter domain.
The proposal step sizes are adapted periodically to target an acceptance rate of approximately \(0.3\), which reduces wasted evaluations from near-certain rejection while maintaining sufficient exploration.

To ensure a like-for-like comparison with \ac{BO}, \ac{SA} uses the exact same adaptive replication evaluator described in Section~\ref{subsec:opt_baseline}, including the same rung schedule and the same uncertainty-aware promotion rule.

\subsubsection{MCMC with Metropolis-Hastings Method}\label{subsubsec:opt_mcmc}

We implement an \ac{MCMC}-style optimizer based on a Metropolis-Hastings acceptance rule, using multiple parallel chains and random-scan proposals that update one parameter at a time.
This design is computationally stable in our setting because each objective evaluation is expensive and the acceptance decision can be made using only the scalar objective estimate.

The acceptance ratio uses the same scaling as Murphy et al. in related work~\cite{murphySynthesizingTuningStochastic2018b}, which treats the score as a probability-like quantity and calibrates acceptance such that a \(1\%\) decrease in score is accepted with probability \(0.25\).
This scaling prevents overly greedy behavior when scores are close to one, while still biasing the search toward high-performing regions.
The proposal scales are adapted periodically to target an acceptance ratio around \(0.25\), which avoids collapse into extremely small steps and reduces wasted evaluations on near-certain rejections.

As for \ac{SA}, the \ac{MCMC} benchmark uses the same adaptive replication evaluator as \ac{BO} and therefore differs only in the search policy and acceptance mechanism.

\subsection{Benchmark Scenarios and Selection Outcome}\label{subsec:opt_selection}

\begin{figure}
    \centering
    \includegraphics[width=0.75\linewidth]{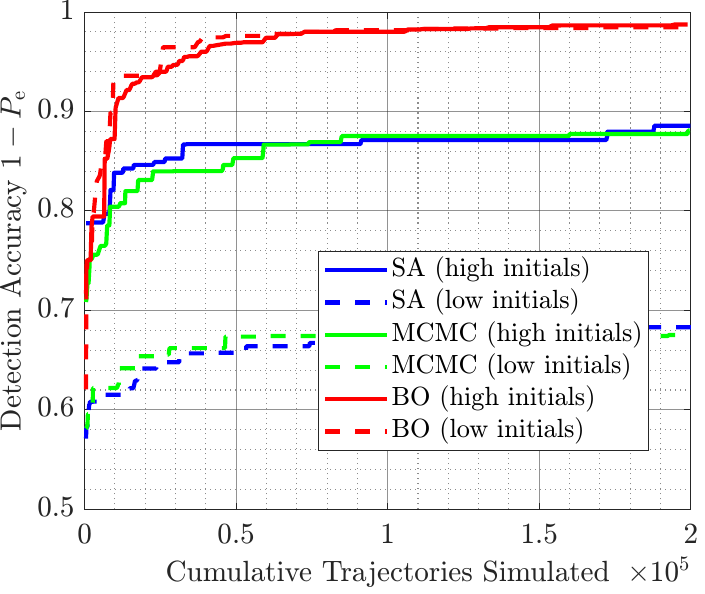}
    \caption{Comparison of optimization schemes showing achieved score versus the cumulative number of simulated \ac{SSA} trajectories on the $x$-axis for \ac{BO}, \ac{SA}, and \ac{MCMC}, using high- and low-value initializations for the \acp{RRC}. Results are shown as an average across 3 independent optimization runs per scheme.}
    \label{fig:opt_comparison}
\end{figure}

We compare \ac{BO}, \ac{SA}, and \ac{MCMC} on representative ChemSICal-Net optimization scenarios, focusing on a fixed processing horizon \(T = T_\mathrm{ref}\).
To incorporate sensitivity to initialization, each scheme is run from two different starting configurations of the \acp{RRC}, namely a high initialization where all \acp{RRC} start near the upper bound and a low initialization where all \acp{RRC} start near the lower bound. For each initialization, three independent optimization runs per scheme are generated.

We summarize the achieved best-so-far score as a function of the cumulative number of simulated trajectories, which provides a fair comparison in terms of computational complexity and resource usage.
Figure~\ref{fig:opt_comparison} shows that \ac{BO} achieves higher scores than both benchmarks for the same simulation budget, reaching accuracies above \(1-P_\mathrm{e} = 0.95\) within approximately \(2\times 10^5\) trajectories and showing consistent performance under both initializations.
In contrast, \ac{SA} and \ac{MCMC} converge to lower scores $<0.9$ and exhibit substantially larger differences between high and low initializations, where the low initialization yields scores $<0.7$.

Based on these results, we select adaptive \ac{BO} as the default optimization scheme for the remainder of the paper because it provides the strongest sample efficiency and the most robust convergence behavior across the tested initializations.

\begin{table*}[t]
\centering\scriptsize
\caption{Simulation and Optimization Parameters}
\label{tab:sim_params_main}
\begin{tabular}{lll}
\hline
Parameter          & Symbol          & Value \\ \hline
Trajectories per input & $N_{\mathrm{traj}}$ & Standard: 200 \\
                       &                     & Optimization Rungs: [20, 60, 100, 500]\\
Optimization rung thresholds & $(1-P_\mathrm{e})_\mathrm{th}$ & [0.9, 0.95, 0.99]\\
Confidence interval & $z$ & $z=1.96$ for 95\% \ac{CI} (proportions) \\
Time reference & $T_\mathrm{ref}$ & 73 [unit-less]\\
Decision time horizon & $T_\mathrm{dec}$ & $\{T_\mathrm{ref}/4, T_\mathrm{ref}/2, T_\mathrm{ref}, 2T_\mathrm{ref}\}$ \\
Input set for evaluation & $Y_\mathrm{on}\in\mathcal{Y}$ &  Standard: $\mathcal{Y} = \{0,1,\dots, 600\}$\\  
& &                                                         Optimization (2-TX): $\mathcal{Y} = \{140, 167, 216, 308, 400, 475\}$\\
& &                                                         Optimization (3-TX): $\mathcal{Y} = \{71, 157, 228, 308, 379, 465, 536\}$\\
Optimization iterations & $N_\mathrm{iter}$ & 1000\\
\hline
\end{tabular}%
\end{table*}

\section{Evaluation of ChemSICal-Net}\label{sec:evaluation}

In the following, we will present a thorough evaluation of the ChemSICal-Net design. Firstly, we will utilize \ac{ODE}-based results of the algorithm performance across the possible \ac{DBMC} channel input range of the $Y_\mathrm{on}$ molecule species. This will provide a first intuition about the more challenging input regions and show the limitations of the \ac{ODE}-based tuning approach, as it leaves a large error probability in the stochastic simulations, see Section~\ref{sec:simulation}.
Then, we systematically apply our \ac{BO} framework to evaluate the stochastic simulation results of the ChemSICal-Net system and optimize the \acp{RRC} and initial molecule concentrations.
We compare different variants of ChemSICal, highlighting the impact of timing control and the \ac{BO} scheme on the error probability $P_\mathrm{e}$. The system with timing control through the chemical oscillator will be denoted as ChemSICal-Net or \textit{Timed}, while the system without timing control is denoted as \textit{Always-On} or \textit{AO}.
Lastly, we present initial evaluations of our concepts for ChemSICal-Net+R, for the reset mechanism, and ChemSICal-Net+TX, for adding another \ac{TX} to the system.
The most important parameters utilized for the communication system, the ChemSICal-Net initial molecule counts and \acp{RRC}, and the \ac{BO} scheme can be found in Tables~\ref{tab:sys_params_main},~\ref{tab:crn_init_main},~\ref{tab:crn_rates_main}, and~\ref{tab:sim_params_main}. Any specific additional parameter values are listed in the supplementary material.

\begin{figure}
    \centering
    \includegraphics[width=0.95\linewidth]{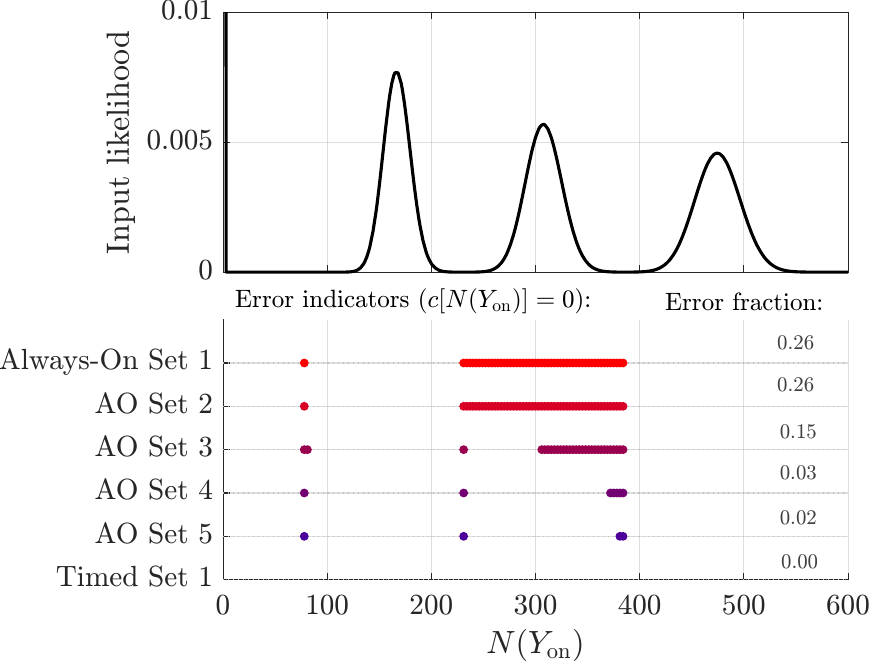}
    \caption{Results of the \ac{ODE}-based evaluation across different \ac{RRC} sets (shown in Table~\ref{tab:ODE_reaction_rates}) for the \textit{Always-On} (AO) and the \textit{Timed} variant of the ChemSICal-Net system. The markers show occurrences of errors within the range of inputs $N(Y_\mathrm{on})$ and the input likelihood is shown on top in black.}
    \label{fig:ode_tuning}
\end{figure}

\begin{table}[t]
\caption{Reaction constant sets utilized in Figure \ref{fig:ode_tuning}.}
\label{tab:ODE_reaction_rates}
\centering
\resizebox{0.9\columnwidth}{!}{%
\begin{tabular}{lllllllll}
\hline
Set & $\kappa_{\mathrm{C},1}$ & $\kappa_{\mathrm{Tr},1}$ & $\kappa_{\mathrm{AM},1}$ & $\kappa_{\mathrm{TA},1}$ & $\kappa_{\mathrm{C},2}$ & $\kappa_{\mathrm{Tr},2}$ & $\kappa_{\mathrm{AM},2}$ \\ \hline
1          & 1.0            & 1.0            & 1.0             & 1.0             & 1.0            & 1.0            & 1.0             \\
2          & 1.0            & 1.0            & 1.0             & 1.0             & 0.1            & 0.1            & 0.1              \\
3          & 1.0            & 1.0            & 0.1             & 1.0             & 0.1            & 0.1            & 0.01            \\
4          & 1.0            & 1.0            & 0.1             & 1.0             & 0.1            & 0.01           & 0.001           \\
5          & 1.0            & 1.0            & 0.1             & 1.0             & 0.1            & 0.1            & 0.001           \\ \hline
\end{tabular}%
}
\end{table}

\subsection{Initial ODE Tuning}

As we identified in prior work~\cite{wietfeldChemSICalEvaluatingStochastic2025}, the \acp{RRC} are the most important factor in determining the correct operation of the ChemSICal-Net algorithm.
Therefore, we will firstly use the fast \ac{ODE} solver to generate an overview of the effect of different sets of \acp{RRC}, as defined in Table~\ref{tab:ODE_reaction_rates}, for both the \textit{Timed} and \textit{Always-On} variant and for a decision time horizon of $T_\mathrm{dec}=T_\mathrm{ref}$. With the \ac{ODE} solver, each input value of $N(Y_\mathrm{on})$ causes a result that is either correct or incorrect. In Figure~\ref{fig:ode_tuning}, the errors ($c[N(Y_\mathrm{on})] = 0$) are indicated in relation to the value of $N(Y_\mathrm{on})$ and the input likelihood distribution $p_\mathrm{n_\mathrm{s}}[Y_\mathrm{on}]$.
For set 1, all \acp{RRC} are set equal to $\kappa_\mathrm{max} = 1$. For the \textit{Always-On} variant, we see a significant number of errors across a large range of input values including ones with high likelihood. This is related to the problem of operation ordering. Since all reactions happen at the same time, if they are also happening at the same speed, later steps in the system are not being computed based on the correct results from prior steps. We show an example of such an issue in Section~\ref{sec:crn_design}, Figure~\ref{fig:trajectory_always_on}. The progression of \ac{RRC} sets 1 to 5 was chosen by hand to showcase how to mitigate this problem in the \textit{Always-On} case. Sets 2 and 3 show that it is not enough to make all reactions of stage-2 slower, or further slowing down the \ac{AM} reactions by an order of magnitude, although set 3 has a slightly reduced error fraction and we are directionally correct with these changes. Lastly, set 4 and 5 suggest that the \ac{AM} reaction for stage-2 must be slower by around three orders of magnitude, removing all but 2\% of the errors in the \ac{ODE} evaluation for the \textit{Always-On} system. This enforces not only ordering between stages but also between different steps within a stage by making some reactions slower than others.
Comparing to the \textit{Timed} system variant, we see that here, the ordering issue is mitigated automatically via the added timing control. Even with set 1, there are no errors indicated by the \ac{ODE} analysis, showing how the \textit{Timed} variant provides simpler initial tuning and setup compared to the \textit{Always-On} variant.
However, as we will show in the following, the actual performance can only be determined through stochastic simulations and calculation of the input-weighted error probability $P_\mathrm{e}$, as explained in Section~\ref{sec:simulation}.

\begin{figure}
    \centering
    \includegraphics[width=0.8\linewidth]{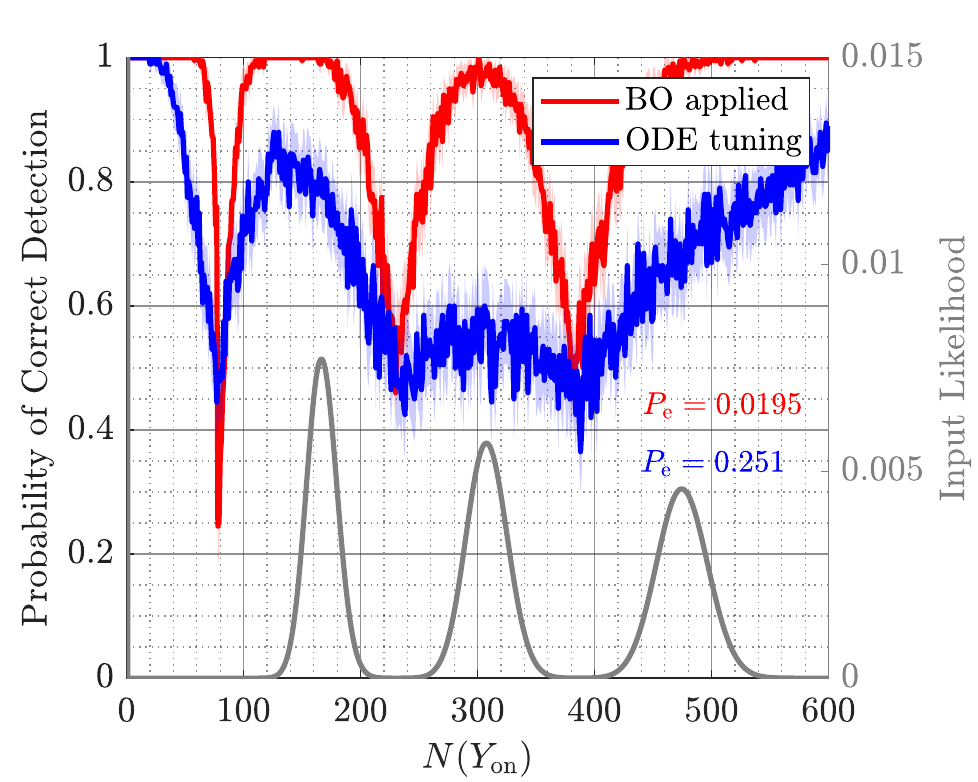}
    \caption{Comparison of the detection accuracy $P_\mathrm{d}[N(Y_\mathrm{on})]$ over the input range of the molecule species $Y_\mathrm{on}$ between the \ac{ODE}-tuned parameters in blue and the \ac{BO}-optimized parameters in red. Input likelihood of $N(Y_\mathrm{on})$ for reference in grey. Input-weighted error probability of the entire system $P_\mathrm{e}$ also shown. \textit{\ac{BO} improves $P_\mathrm{d}$ a lot in high-likelihood regimes; $P_\mathrm{e}$ is reduced from 0.25 to 0.02.}}
    \label{fig:basic_case}
\end{figure}

\subsection{Stochastic Simulation and Impact of Bayesian Optimization}

The parameters for stochastic evaluation used for all results shown in the following can be found in Table~\ref{tab:sim_params_main}.
Figure~\ref{fig:basic_case} shows the resulting probability of correct detection $P_\mathrm{d}[N(Y_\mathrm{on})]$ across the range of possible input values on the $x$-axis, as well as the input likelihood for reference in grey. The thick colored lines show the average value, while the shaded area is associated with the 95\% \ac{CI}.
The blue line depicts $P_\mathrm{d}$ for the \textit{Timed} variant using set 1, identified as error-free during the \ac{ODE} tuning described above in Figure~\ref{fig:ode_tuning}. The results show that while the average value of $P_\mathrm{d}$ largely remains above 50\%, as indicated by the \ac{ODE} analysis, there are large parts of the input space with $P_\mathrm{d}$ at or close to 50\%. The input-weighted error probability is $P_\mathrm{e} = 0.25$. This means that in this configuration ChemSICal-Net does not align with the desired algorithm behavior in many cases that occur with high likelihood.
The red line in Figure~\ref{fig:basic_case} depicts $P_\mathrm{d}[N(Y_\mathrm{on})]$ after we have applied our proposed adaptive \ac{BO} scheme, using the setup and parameters described in Section~\ref{sec:optimization}. We can see that the performance can be significantly increased across the entire range of input values with a particular emphasis on the most likely regions. $P_\mathrm{d}$ retains the characteristic shape with peaks and minima roughly in the same places where the input likelihood has peaks and minima. This is caused by the correct choice of threshold concentrations and is very advantageous towards reducing the input-weighted error probability, since the input regions with the lowest $P_\mathrm{d}$ also have the lowest likelihood. Overall, the application of the \ac{BO} scheme reduces $P_\mathrm{e}$ from 0.25 to $\approx 0.02$.

\begin{figure}
    \centering
    \includegraphics[width=0.75\linewidth]{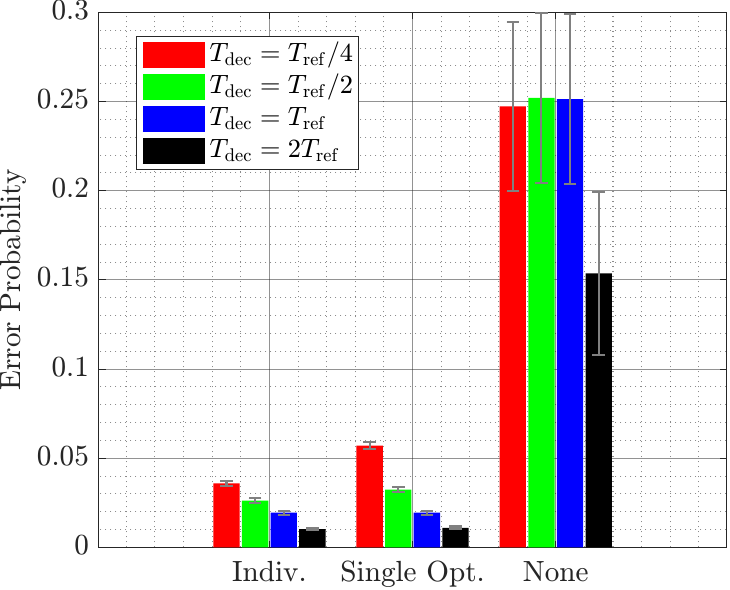}
    \caption{Comparison of the input-weighted error probability $P_\mathrm{e}$ for different decision time horizons $T_\mathrm{dec}$ and different optimization configurations. For \textit{Indiv. Opt.}, \ac{BO} scheme is run for each value of $T_\mathrm{dec}$ individually. For \textit{Single Opt.}, \ac{BO} is only run for $T_\mathrm{dec}=T_\mathrm{ref}$ and results applied to all other values of $T_\mathrm{dec}$. For \textit{None}, parameters correspond to set 1 in Table~\ref{tab:ODE_reaction_rates}. \textit{\ac{BO} improves $P_\mathrm{e}$ across all $T_\mathrm{dec}$; individual tuning matters most for short $T_\mathrm{dec}$.}}
    \label{fig:opt_bar_comparison}
\end{figure}

To investigate the impact of optimization further, we incorporate different values of the decision time horizon $T_\mathrm{dec}$ and two different optimization configurations. We denote the case, where we individually run the \ac{BO} scheme for each value of $T_\mathrm{dec}$ as \textit{Indiv. Opt.}. We further perform an analysis, where we only optimize the parameters for $T_\mathrm{dec}=T_\mathrm{ref}$ once, and then apply those same parameters for all other values of $T_\mathrm{dec}$. This is denoted as \textit{Single Opt.}. When we use the parameters from \ac{ODE} tuning, we denote this optimization configuration as \textit{None}.
The results are shown in Figure~\ref{fig:opt_bar_comparison} with $P_\mathrm{e}$ on the $y$-axis and the values depicted as bars with a 95\% \ac{CI}. We can see that \textit{Indiv. Opt.} with the \ac{BO} scheme improves $P_\mathrm{e}$ by about one order of magnitude across all values of $T_\mathrm{dec}$. Between the different decision times, i.e. larger $T_\mathrm{dec}$ means more time to decide, we can see that $P_\mathrm{e}$ and $T_\mathrm{dec}$ are directly connected. The lowest error probability is achieved, when ChemSICal-Net has the largest amount of decision time $T_\mathrm{dec}=2T_\mathrm{ref}$, creating a clear trade-off between detection accuracy and decision time.
Comparing between \textit{Indiv. Opt.} and \textit{Single Opt.}, we can see that tailoring the parameters specifically for one value of $T_\mathrm{dec}$ improves the performance further. However, the benefit becomes larger for shorter decision times $T_\mathrm{dec}$.

\subsection{Impact of Timing Control}

\begin{figure}
    \centering
    \includegraphics[width=0.6\linewidth]{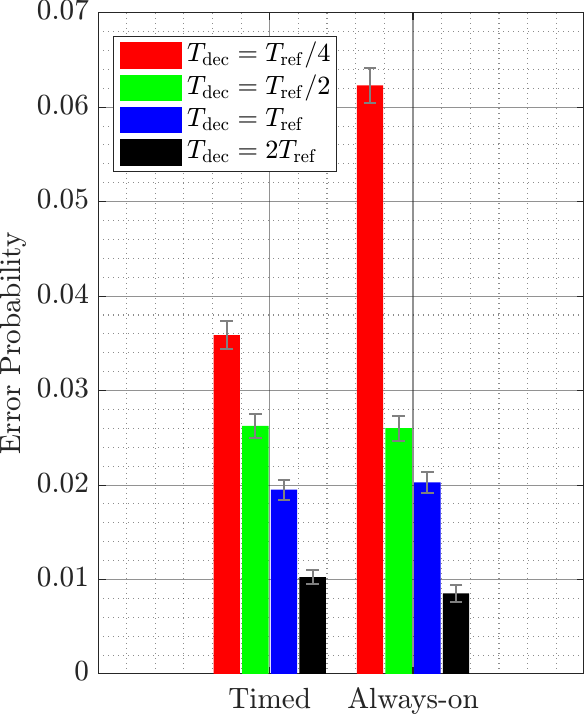}
    \caption{Comparison of the input-weighted error probability $P_\mathrm{e}$ for different decision time horizons $T_\mathrm{dec}$ and two variants of ChemSICal-Net. For \textit{Timed}, the chemical clock is used to gate the different stages. For \textit{Always-On}, no clock gating is applied. \textit{Timing helps for short $T_\mathrm{dec}$; for longer horizons, less overhead is preferable.}}
    \label{fig:timed_comparison}
\end{figure}

In Figure~\ref{fig:timed_comparison}, comparison results are presented for the stochastic evaluation of the \textit{Timed} and \textit{Always-On} variants after applying the \ac{BO} scheme to both. The initial parameters for the optimization are defined by the respective results of the \ac{ODE} tuning above.
Across the different values of $T_\mathrm{dec}$ a trend emerges. We can observe a middle region, i.e. $T_\mathrm{dec}=T_\mathrm{ref}$, and $T_\mathrm{dec}=T_\mathrm{ref}/2$, where both variants perform relatively similar with a slight advantage for the \textit{Timed} variant but within the \ac{CI}.
For the fast-decision case, i.e. $T_\mathrm{dec}=T_\mathrm{ref}/4$, the \textit{Timed} variant clearly outperforms the \textit{Always-On} variant by approximately a factor of 2. This demonstrates the ability of the timing control to separate computing steps in faster-paced environments.
In contrast, for long decision times, i.e. $T_\mathrm{dec}=2T_\mathrm{ref}$, the results show a small but visible advantage for the \textit{Always-On} scheme by a factor of 1.2. In this regime, the ChemSICal-Net system has a long time for any molecule species to converge to correct values and also correct initial wrong decisions over time, so that the issues with premature decisions outlined in Section~\ref{sec:crn_design} are not as relevant. At the same time, the addition of more reactions and gating effects in the \textit{Timed} variant cause more variance and distortion to the intended reaction block operation. This causes a slightly higher error rate.

\subsection{Relation to the Analytical System}
In our previous work~\cite{wietfeldDBMCNOMAEvaluatingNOMA2024, wietfeldDBMCaNOMAlyAsynchronousNOMA2025}, we have proposed and analyzed a non-chemical version of the \ac{SIC} algorithm for \ac{DBMC-NOMA}. The channel model and communication system assumptions were identical to the ones utilized in this paper so that the results are comparable. In the computationally accurate implementation without \acp{CRN}, the derived analytical model predicts an error probability of approximately $10^{-6}$, i.e. four orders of magnitude smaller than the smallest reported $P_\mathrm{e}$ by the ChemSICal-Net system for the longest decision time $T_\mathrm{dec}$. This highlights the fact that the chemical implementation of an algorithm designed for \ac{DBMC} systems will most likely dominate the error characteristic of the system over the inherent error generated by the channel noise and modulation and detection scheme~\cite{wietfeldChemSICalEvaluatingStochastic2025}. As a result, an improvement in the error probability of the chemical implementation will directly translate into performance gains of the entire system and, therefore, the consideration of the chemical aspects should be prioritized. Additionally, the propagation and computation time will similarly be dominated by the chemical reaction dynamics. Although we use normalized reaction speeds in this work, realistic time scales for appropriate reactions like enzyme- or DNA strand-based systems are on the order of minutes~\cite{wietfeldModelingMicrofluidicInteraction2025b, kholodenkoNegativeFeedbackUltrasensitivity2000}, while diffusion transport over micro- or nanoscale distances, or flow-based transport across larger distances can take below one second to several seconds~\cite{wietfeldEvaluationMultiMoleculeMolecular2024b}. Therefore, reducing the decision time of the \ac{CRN} system by appropriate tuning of the reaction parameters will be crucial.

\subsection{Proof-of-Concept Extensions}

Lastly, we want to highlight the results of our preliminary extensions of the ChemSICal-Net system, which showcase the potential for scaling and implementation as well as future challenges that need to be addressed.

\subsubsection{Reset Mechanism}

\begin{figure}
    \centering
    \includegraphics[width=0.75\linewidth]{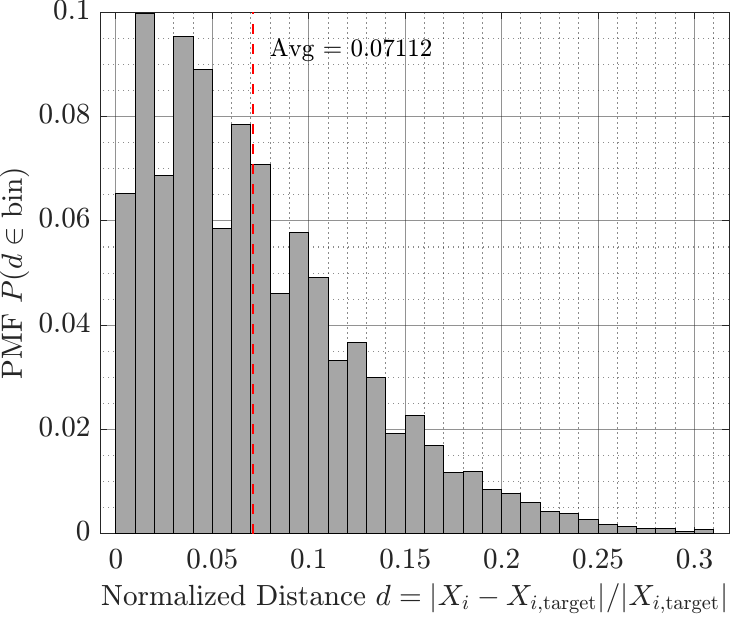}
    \caption{Plot evaluating the success of the reset mechanism. The figure shows a binned probability mass function of the relative normalized distance of the $X$ comparison species from its respective target value after reset.}
    \label{fig:reset_results}
\end{figure}

For our evaluation of the ChemSICal-Net+R reset mechanism, we performed stochastic simulations of the standard ChemSICal-Net reaction network across the entire input range and added the reset phase at the end. Thereby, the molecule species ended up in various possible constellations before having to be reset to the target values. Additionally, we applied our \ac{BO} scheme to optimize the parameters of the reset mechanism. The results showed that the \acp{RRC} controlling the individual reset reactions should be set to $\kappa_\mathrm{max} = 1$, i.e. as fast as possible. In any case, due to the simplicity of the reactions, the reset process takes much less time than the \ac{SIC} decision time, with $T_\mathrm{R} \ll T_\mathrm{dec}$. 

The decision species $D_i^0$ and $D_i^1$, as well as other intermediate species must return to zero molecule count. In our evaluation, this was successful in all simulated trajectories. The intermediate species $X_{\mathrm{off},i}$ and $X_{\mathrm{on},i}$ encode the threshold adaptation values for the \ac{SIC} decision tree and need to return to their original non-zero values. In Figure~\ref{fig:reset_results}, we show a binned probability mass function over the relative normalized distances of all $X$ species from their target at the end of the reset period $T_\mathrm{R}$. We can see that the species count deviated on average by about 7\% from the targets, while 95\% of species returned to within 20\% of their target. The reset mechanism, therefore, works in principle, but should be tuned and evaluated further in future work.

\subsubsection{Adding a Third TX}

\begin{figure}
    \centering
    \includegraphics[width=0.85\linewidth]{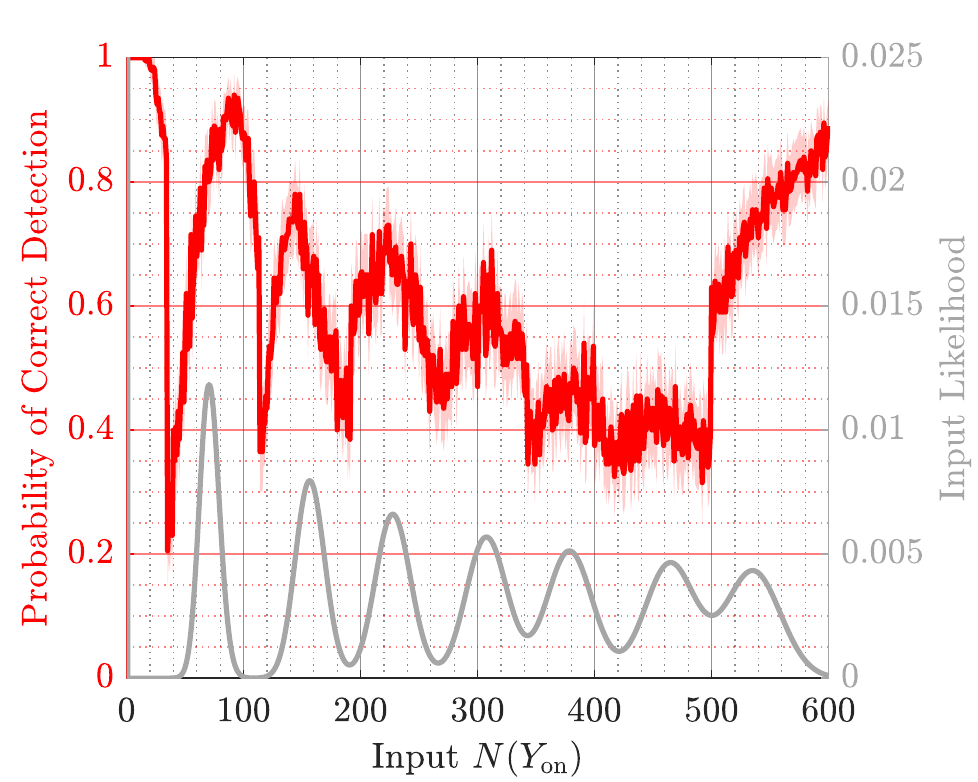}
    \caption{Plot of $P_\mathrm{d}$ in red for the extension with an additional \ac{TX}. Input likelihood for the 3-\ac{TX} case is shown in grey.}
    \label{fig:3tx_results}
\end{figure}

For this evaluation, we set up the system as described in Section~\ref{subsubsec:crn_design_additional_txs} to add another \ac{TX} to the \ac{SIC} decoding process. We run stochastic simulations with the standard parameters, as in Table~\ref{tab:sim_params_main} and apply our \ac{BO} scheme to optimize the \ac{CRN} parameters. Here, we only looked at the case where $T_\mathrm{osc}=2T_\mathrm{ref}$. For the 3-TX variant we set $T_\mathrm{osc} = \frac{3}{2}T_\mathrm{dec}$, so that three half-cycles of the oscillation are contained within $T_\mathrm{dec}$ to represent the three stages. Figure~\ref{fig:3tx_results} depicts the resulting $P_\mathrm{d}$ and the input likelihood. It is clearly visible that the input likelihood $p_{n_\mathrm{s}}[N(Y_\mathrm{on})]$ is different from the 2-TX case due to the larger number of possible symbol combinations of the three \acp{TX}. The plotted $P_\mathrm{d}$ exhibits the same characteristic shape as before, where the peaks and minima line up with the likelihood. However, in this case, the input likelihood does not have the same pronounced minima as the 2-\ac{TX} system, making it difficult to avoid the strong influences of dips in the detection accuracy between two peaks and causing a lower accuracy overall. We observe a significant symbol-combination dependence of the detection accuracy, where symbols with higher arriving molecule count result in a lower accuracy. This scenario reveals one of the drawbacks of the efficiency-focused input sampling for the optimization scheme. As we are only considering the likelihood peaks for the optimization, the minima are not considered. However, in the final evaluation they have a much larger influence due to the shape of the input likelihood in this case. In the future, the optimization should be tuned to take into account more values close to the midpoints between peaks given that there is more time and computing power available.
In general, the mechanism to expand the number of \acp{TX} is functional and structurally sound and exhibits sub-exponential complexity scaling as explained in Section~\ref{subsubsec:crn_design_additional_txs}. Future evaluations should look into the molecule counts necessary to make larger numbers of \acp{TX} viable with higher detection accuracy.

\section{Conclusion and Further Work}\label{sec:conclusion}

This work presented ChemSICal-Net, an end-to-end chemical \ac{RX} framework that realizes \ac{SIC}-based multi-user reception for \ac{DBMC} \ac{MA} using a stochastic \ac{CRN}. We mapped the algorithmic \ac{SIC} pipeline to modular reaction blocks, introduced oscillator-based timing control to gate multi-stage processing, and integrated an adaptive \ac{BO} scheme to tune constrained \acp{RRC} with communication-centric objectives that improved error probability from the non-optimized baseline by one order of magnitude. Performance was evaluated using deterministic \ac{ODE} analysis and \ac{MCS} based on stochastic simulation trajectories, highlighting how timing control reduced error probability by 2x for short decision time horizons, and showing that a non-clocked design can be beneficial due to lower reaction overhead. In addition, we provided an explicit benchmarking comparison across oscillator model families and optimization schemes, which clarifies and motivates the design choices made for ChemSICal-Net.

Key limitations concern the overall \ac{CRN} complexity and the accuracy and portability of kinetic parameters, the currently abstracted channel-to-\ac{CRN} interface, i.e. sample-and-hold of the received observation, and the practical feasibility and time-scale of the oscillator implementation. Next, we will investigate implementations of the proposed chemical circuits using DNA strand displacement, leveraging our microfluidic interaction-channel platform~\cite{wietfeldModelingMicrofluidicInteraction2025b}, and extend the framework to larger numbers of \acp{TX} by expanding the control mechanism through timing and multi-species oscillator designs.




\bibliographystyle{IEEEtran}
\bibliography{references}


 




\vfill

\end{document}